\documentclass[12pt]{iopart}
\usepackage{epsfig,float,afterpage,amssymb}
\newcommand{\be}{\begin{equation}}
\newcommand{\ee}{\end{equation}}
\newcommand{\bea}{\begin{eqnarray}}
\newcommand{\eea}{\end{eqnarray}}
\newcommand{\w}{\omega}
\newcommand{\bra}{\langle}
\newcommand{\cc}{\mbox{\scriptsize{c}}}
\newcommand{\m}{\mbox{\scriptsize{m}}}
\newcommand{\ket}{\rangle}
\newcommand{\subi}{\mbox{\scriptsize{I}}}
\newcommand{\subr}{\mbox{\scriptsize{R}}}
\newcommand{\band}{\mbox{\scriptsize{b}}}
\newcommand{\delr}{\Delta_0^{\frac{1}{1-r}}}
\newcommand{\up}{\uparrow}
\newcommand{\down}{\downarrow}
\newcommand{\ra}{\rightarrow}
\newcommand{\ssz}{\scriptsize}
\newcounter{saveeqn}
\newcommand{\alpheqn}{\setcounter{saveeqn}{\value{equation}}%
\setcounter{equation}{0}%
\addtocounter{saveeqn}{1}%
\renewcommand{\theequation}{\mbox{\arabic{section}.\arabic{saveeqn}\alph{equation}}}%
}
\newcommand{\appalpheqn}{\setcounter{saveeqn}{\value{equation}}%
\setcounter{equation}{0}%
\addtocounter{saveeqn}{1}%
\renewcommand{\theequation}{\mbox{\Alph{section}.\arabic{saveeqn}\alph{equation}}}%
}
\newcommand{\app}{\setcounter{equation}{0}%
\setcounter{section}{1}%
\renewcommand{\thesection}{\Alph{section}}%  
\renewcommand{\theequation}{\thesection.\arabic{equation}}}%

\newcommand{\reseteqn}{\setcounter{equation}{\value{saveeqn}}%
\renewcommand{\theequation}{\arabic{section}.\arabic{equation}}}
\newcommand{\appreseteqn}{\setcounter{equation}{\value{saveeqn}}%
\renewcommand{\theequation}{\Alph{section}.\arabic{equation}}}
\newcommand{\seceq}{\setcounter{equation}{0}}

\begin{document}
\jl{31}
%\submitted
\title{A Local Moment Approach to magnetic impurities in gapless Fermi systems}
\author{David E Logan and Matthew T Glossop}
\address{Oxford University, Physical and Theoretical Chemistry Laboratory, South Parks Road, Oxford OX1 3QZ, UK}
\tolerance=50
\begin{abstract}
 A local moment approach is developed for single-particle
excitations of a symmetric Anderson impurity model (AIM) with a
soft-gap hybridization vanishing at the Fermi level:
$\Delta_{\subi} \propto |\omega|^{r}$, with $r>0$. Local moments are
introduced explicitly from the outset, and a two-self-energy
description is employed in which single-particle excitations are
coupled dynamically to low-energy transverse spin fluctuations.
The resultant theory is applicable on all energy scales, and 
captures both the spin-fluctuation regime of strong coupling
(large-$U$), as well as the weak coupling regime where it is perturbatively
exact for those $r$-domains in which perturbation theory in $U$ is
non-singular. While the primary emphasis is on single-particle dynamics,
the quantum phase transition between strong coupling (SC) and local
moment (LM) phases can also be addressed directly; for the spin-fluctuation
regime in particular a number of asymptotically exact results are thereby obtained,
notably for the behaviour of the critical $U_{\cc}(r)$ separating SC/LM states and the
Kondo scale $\omega_{\m}(r)$ characteristic of the SC phase. Results
for both single-particle spectra and SC/LM phase boundaries are found to
agree well with recent numerical renormalization group (NRG) studies; and a
number of further testable predictions are made. Single-particle
spectra are examined systematically for both SC and LM states; in particular,
for \it all \rm $0 \leq r < \frac{1}{2}$, spectra characteristic of the SC state are predicted
to exhibit an $r$-dependent universal scaling form as the SC/LM phase
boundary is approached and the Kondo scale vanishes. Results for the
`normal' $r=0$ AIM are moreover recovered smoothly from the limit $r\rightarrow 0$,
where the resultant description of single-particle dynamics includes recovery of
Doniach-\u{S}unji\'{c} tails in the wings of the Kondo resonance, as well as characteristic
low-energy Fermi liquid behaviour and the exponential diminution with $U$ of the
Kondo scale itself. The normal AIM is found to represent a particular case
of more generic behaviour characteristic of the $r>0$ SC phase which, in agreement
with conclusions drawn from recent NRG work, may be viewed as a non-trivial but natural generalization
of Fermi liquid physics.

\end{abstract}
\maketitle
%\begin{spacing}{1.5}
\seceq
\section{Introduction}
  The Anderson impurity model (AIM) [1] is the archetype for describing
dilute, correlated magnetic impurities in metals. Reviewed
comprehensively in [2], its essential strong coupling behaviour is 
that of the Kondo effect: the spin-$\frac{1}{2}$ impurity is quenched by coupling
to low-energy excitations of the non-interacting metallic host. But
while thermodynamic properties of the model are well understood, the
same cannot be said for a theoretical description of dynamics, in particular
those of single-particle excitations. Here a wide variety of theories have been
developed, including the non-crossing approximation (NCA) [3-6], $1/N$ expansions 
[7-9] and slave boson approaches [10-12]. Their undoubted successes notwithstanding,
however, each has significant limitations. They are designed to capture
the $N \rightarrow \infty $ limit (as opposed to the spin-$\frac{1}{2}$ case, $N=2$),
and the extreme asymmetric limit of $U=\infty$ with $U$ the local (impurity)
interaction: extension to finite-$U$ is not straightforward. The NCA 
describes well high-energy single-particle excitations embodied in the
Hubbard satellite bands, but fails to capture the low-energy Fermi
liquid behaviour of the Kondo resonance. Slave boson approaches by
contrast are much less satisfactory on high-energy scales, their virtue
being to handle low-energies and Fermi liquid behaviour. Even here
however they are limited: in common with $1/N$ expansions, they cannot
for example recover the Doniach-\u{S}unji\'{c} tails [13] in the Kondo resonance
that are observed in numerical renormalization group [14] and quantum
Monte Carlo [15] studies of the AIM, and known to be important
experimentally [16].

  The normal AIM has of course one `simplifying' feature: the host is 
metallic by presumption, whence the host-impurity coupling embodied
in the hybridization function $\Delta(\omega)$ is essentially 
frequency-independent and controlled by its value at the Fermi level,
$\omega=0$; in consequence, and excepting the atomic limit where the
impurity/host trivially decouple, the impurity spin is quenched and
the system is a conventional Fermi liquid for all $U\geq0$ (see e.g. [2]).
The brief remarks above are nonetheless not confined to this problem,
but symptomatic in general of the widely accepted need for new
theoretical approaches to strongly correlated electrons. And the 
challenge is naturally more acute when $\Delta(\omega)$ acquires
a frequency dependence that can lead to qualitatively new physics, and
in particular the possibility of nontrivial zero-temperature phase
transitions. One example of such arises in the lattice-based Hubbard model,
which within the framework of Dynamical Mean-Field Theory (or the 
infinite-dimensional limit) maps onto an effective AIM [17] with a hybridization
that is a functional of the impurity Green function, and is hence both
$\omega$-dependent and to be determined self-consistently. The quantum phase
transition here is the celebrated and still controversial Mott transition,
occurring at a critical $U_{\cc}$ in the paramagnetic phase of the half-filled
(particle-hole symmetric) model: between a gapless Fermi liquid metal, and
a gapped local moment insulator characterized by an $k_{\mbox{\ssz{B}}}\mbox{ln}2$ residual entropy
(for reviews, see e.g. [18-20]).

  A second example, considered in this paper, is the soft-gap AIM
appropriate to a semi-metallic host, or one that itself may be viewed as
being on the verge of a simple
band-crossing metal-insulator transition: in which the (imaginary part of the)
hybridization function exhibits a soft-gap at the Fermi level,
$\Delta_{\subi}(\omega)\propto|\omega|^{r}$ with $r>0$, in contrast to
the normal `metallic' AIM, $r=0$, for which $\Delta_{\subi}(\omega=0)$ is constant.
This problem, for which a wide range of possible experimental
candidates arise (see eg [21]),  was first studied by Withoff and Fradkin [22] in
the context of the corresponding soft-gap Kondo model, using both `poor man's'
scaling and a large-$N$ mean-field theory. Much study of the soft-gap Kondo
and Anderson models has since ensued, in particular via scaling [22,23], large-$N$
expansions [22,24,25], the numerical renormalization group (NRG) [21,23,26-28] and
perturbation theory in $U$ [29]. It is known thereby that two distinct ground states
exist, between which in general a quantum phase transition occurs at a finite
critical $U_{\cc}(r)$: a doubly degenerate local moment (LM) state in which the
impurity spin remains unquenched; and a strong coupling (SC) state in which the
impurity spin is locally quenched, and a Kondo effect is manifest.

 The underlying physics is known to be particularly rich for the particle-hole
symmetric model, to which NRG studies in particular have devoted considerable
attention, including both thermodynamic properties [21,26-28] and impurity
single-particle spectra [28]. It is the symmetric spin-$\frac{1}{2}$ soft-gap AIM that we 
consider here, by developing a microscopic `local moment approach' that has recently
been applied successfully to the normal ($r=0$) AIM [30]. Our primary focus is thus
an analytical treatment of single-particle dynamics --- on all energy scales, and for
any interaction strength $U$ --- as embodied in the impurity Green function,
$G(\omega)$, and hence spectrum $D(\omega)\propto \mbox{Im}G(\omega)$; although an
integral element of the approach also permits statics, in the form of the
SC/LM transition and associated phase boundaries, to be addressed directly.

  The resultant theory, which seeks in particular to capture the spin-fluctuation
regime of strong coupling (large-$U$) but is also perturbatively exact
in weak coupling for those $r$-domains in which straight perturbation theory
in $U$ is known to be applicable [29], appears to be rather successful. Its predictions
for single-particle spectra and SC/LM phase boundaries agree well, both qualitatively
and quantitatively, with extant NRG calculations [21,28] that provide essentially
exact numerical results for the problem. In the spin-fluctuation regime of strong
coupling a number of asymptotically exact results are obtained, in particular for
the Kondo scale characteristic of the SC phase and the low-$r$ behaviour of the
critical $U_{\cc}(r)$. Results for the normal AIM [30] are moreover obtained smoothly
from the limit $r\rightarrow0$, where the resultant description of single-particle
dynamics includes recovery of the Doniach-\u{S}unji\'{c} tails, as well as
characteristic low-$\omega$ Fermi liquid behaviour and the exponentiality
of the Kondo scale. We find in fact that the normal AIM constitutes in many
ways just a particular case of more generic behaviour characteristic of
the $r>0$ SC state, which we argue may be regarded as a `generalized Fermi
liquid', in agreement with the conclusions drawn by Gonzalez-Buxton and
Ingersent [21] from detailed NRG studies. One manifestation of this is the
prediction that, for all $r\in[0,\frac{1}{2})$, single-particle spectra characteristic
of the SC state should acquire a universal scaling form as the SC/LM phase
boundary is approached, thus generalizing to finite-$r$ behaviour that is
familiar in the context of the normal AIM [14,15]; this and related predictions of the
present theory will be tested against NRG calculations in a subsequent
publication [31].

  The paper is organized as follows. A brief introduction to the soft-gap
AIM is given in section 2, where two facets are highlighted. First, the
non-interacting limit $U=0$. Its behaviour, in contrast to that
of the normal AIM, is nontrivial: both SC and LM states arise, for
$r<1$ and $r>1$ respectively, and with distinct signatures in the underlying
spectral functions [29]. Second, we emphasise the generalized pinning condition
established by us previously [29], whereby $A(\omega)=|\omega|^{r}D(\omega)$ is
pinned at the Fermi level $\omega=0$ for any $r$ and $U$ where a SC state
obtains; and which represents a generalization of the corresponding condition
familiar for the $r=0$ AIM where it is normally viewed as a consequence of
the Friedel sum rule (see e.g. [2]). Imposition of this spectral pinning as
a self-consistency condition plays a central role in the current work.

  In section 3 we introduce the `two-self-energy' description that underlies 
the present theory. Such an approach is physically natural if one
aims to describe the doubly degenerate LM state; and, we would argue,
is at least desirable if one seeks to construct
a non-perturbative theory that can simultaneously handle the possibility
of both LM and SC states, and hence the transition between
them. As for the normal AIM [30], our approach to the interaction self-energies
starts from the simplest non-trivial mean-field approximation in which the
notion of an impurity local moment, determined self-consistently, is introduced
explicitly from the outset: unrestricted Hartree-Fock (UHF), as considered by
Anderson in his original paper [1] on the $r=0$ AIM. The deficiencies of this
static mean-field approximation {\it per se} are of course severe; but it is
in large part a physical understanding of them, considered in section 4.2, that enables 
a subsequent many-body approach to be developed successfully. Moreover, and in
contrast to the normal AIM, even UHF is nontrivial for the soft-gap problem: as
shown in section 4.1 it gives rise for example to a phase diagram that, in predicting
for all finite-$U$ {\it solely}  LM states for any $r>\frac{1}{2}$, concurs qualitatively
with the results of NRG calculations [21,28]. 

  Dynamical many-body contributions to the self-energies, over and above the 
Fock term alone retained at simple mean-field level, are detailed
in section 5 for both SC and LM states. In physical terms these embody coupling
of single-particle excitations to low-energy transverse spin fluctuations,
and capture the dynamical spin-flip scattering required in particular to
describe the Kondo, or spin-fluctuation, regime. SC states are obtained {\it via}
self-consistent imposition of the generalized pinning condition, the limits
of stability of solutions to which in turn give the critical $U_{\cc}(r)$ for the
SC/LM transition. In this way, as for the $r=0$ AIM [30], the Kondo or spin-flip
scale $\omega_{\m}\equiv\omega_{\m}(r)$ arises naturally within the present
approach, and is found to be non-zero throughout the entire SC phase, vanishing
continuously only as the SC/LM phase boundary is approached, $U\rightarrow U_{\cc}(r)-$.
For $U>U_{\cc}(r)$ in the LM phase, by contrast, $\omega_{\m}(r)=0$: as expected
physically for a doubly degenerate state, where there is no energy cost for
a local spin flip.

  Resultant phase boundaries between SC and LM states are considered
in section 6; including (section 6.3) their predicted one-parameter scaling behaviour in
the regimes $U/D\gg 1$ and $\ll 1$ (with $D$ the bandwidth of the
host spectrum or hybridization function), and detailed comparison
with extant NRG results [21,28] (section 6.4). The evolution and critical behaviour of the
SC Kondo scale $\omega_{\m}(r)$ is considered in section 6.1. Particular emphasis is given
here to small $r\ll 1$ where salient results can be extracted analytically;
and which is of evident importance in connecting to the normal $r=0$ AIM, 
for which the exact exponential asymptotics of the Kondo scale are correctly
recovered. As $r\rightarrow 0$, the critical $U_{\cc}(r)$ for the SC/LM boundary
is found to be $\propto 1/r$, and the condition for the Schrieffer-Wolff
transformation [32] mapping the soft-gap AIM to the corresponding Kondo model is
thus satisfied. The critical exchange coupling $J_{\cc}(r)$ for the soft-gap
Kondo model as $r\rightarrow 0$ can thus be obtained from the present approach
(section 6.2); and is found to be given precisely by the scaling result obtained originally
by Withoff and Fradkin [22], which we argue is asymptotically exact.

  Single-particle impurity spectra, and their evolution with interaction strength
$U$ from strong to weak coupling, are considered explicitly in sections 7 and 8. The
`bare' $D(\omega)$ are discussed in section 7, on all energy scales and for
both SC and LM states. Many-body broadening of the high-energy
Hubbard satellites, whose existence is well known for the normal AIM (see eg [6,30]),
is argued to arise also in the soft-gap problem and shown to be correctly
recovered by the present approach; as too are the characteristic 
$\omega\rightarrow 0$ spectral signatures of the SC and LM phases found in
NRG calculations [28], viz $D(\omega)\sim |\omega|^{-r}$ and $\sim |\omega|^{r}$
respectively. The behaviour of the spectra in weak coupling, $U\rightarrow 0$,
is considered in section 7.1. For $r<\frac{1}{2}$ where both the $U\rightarrow 0$ and the $U=0$
ground states are SC, {\it and}  for $r>1$ where the ground state is found to be
LM for all $U\geq 0$, the theory is shown to be perturbatively exact to
(and including) second-order in $U$ about the noninteracting limit. For 
$\frac{1}{2}<r<1$ by contrast, the $U>0$ ground state is found to be LM but the non-interacting 
ground state is SC [29]; and the natural breakdown of conventional perturbation 
theory in $U$ is clearly evident in the non-analyticity (in $U$) of the conventional 
`single' self-energy as $U\rightarrow 0$. Explicit comparison to single-particle 
spectra obtained from NRG calculations [28] is made in section 7.2, and excellent
agreement is found. 

  Finally, we consider in section 8 the spectral functions $A(\omega)=|\omega|^{r}D(\omega)$
in the SC phase --- `modified' to remove the $|\omega|^{-r}$ divergence at 
low-$\omega$ that is symptomatic of the SC state, and which is entirely
unrenormalized by interaction effects [29]. The $A(\omega)$ are
found to exhibit familiar characteristics: they are pinned at the Fermi level,
$\omega =0$, and contain a generalized Kondo resonance whose width is proportional
to the Kondo scale $\omega_{\m}(r)$ and thus narrows progressively as $U$ is 
increased towards the SC/LM phase where $\omega_{\m}(r)$ vanishes. This is
just as found for the normal AIM (see eg [2,14,15]) where the critical
$(U/\Delta_0)_{r=0}=\infty$  (reflecting the fact that the SC/LM transition here
coincides trivially with the atomic limit). Moreover, as for the $r=0$ AIM,
we find generally for {\it any}  given $r<\frac{1}{2}$ that as $U\rightarrow U_{\cc}(r)-$,
the $A(\omega)$ becomes a universal function of $\omega/\omega_{\m}(r)$; with
a scaled Kondo resonance that exhibits characteristic $r$-dependent low-frequency
behaviour, as well as Doniach-\u{S}unji\'{c} tails in the `wings' of the spectrum
for $\omega/\omega_{\m}(r)\gtrsim 1$.

\seceq
\section{Background}
We begin with some necessary background, particularly in relation to the non-interacting limit (section 2.1), and the generalized spectral pinning condition characteristic of a SC state with $U>0$ [29] (section 2.2).

With the Fermi level taken as the energy origin, the Hamiltonian for the spin-$\frac{1}{2}$ Anderson model is given in standard notation by
\alpheqn
\reseteqn
\setcounter{equation}{0}
\begin{eqnarray}
\hat{H}& = \hat{H}_{\mbox{\ssz{host}}}+\hat{H}_{\mbox{\ssz{impurity}}}+\hat{H}_{\mbox{\ssz{hybridization}}}\nonumber \\
& = \sum_{\bi{k}, \sigma}\epsilon_{\bi{k}}\hat{n}_{\bi{k} \sigma}+\sum_{\sigma}(\epsilon_{i}+\mbox{$\frac{1}{2}$}U\hat{n}_{i-\sigma})\hat{n}_{i\sigma}+\sum_{\bi{k},\sigma}V_{i\bi{k}}(c^{\dagger}_{i\sigma}c_{\bi{k}\sigma}+c^{\dagger}_{\bi{k}\sigma}c_{i\sigma})
\eea
with $\epsilon_{\bi{k}}$ the host dispersion, $V_{i\bi{k}}$ the hybridization and $\epsilon_{i}$ the impurity level; for the symmetric Anderson model considered here, $\epsilon_{i}=-U/2$ with $U$ the on-site interaction.

We focus on the zero-temperature single-particle impurity Green function, defined by
\be 
G(t) = -\mbox{i}\bra T\{c_{i\sigma}(t)c^{\dagger}_{i\sigma}\}\ket = G^+(t)+G^-(t)
\ee
and separated for later purposes into retarded and advanced components; since $\hat{H}$ is invariant under $\sigma \rightarrow -\sigma$, $G$ is naturally independent of spin, $\sigma$.  We add that while the primary physical content of $G(\omega)$ is that of single particle dynamics, analysis of it will also enable identification of the phase boundaries between SC and LM states, as detailed in sections 5 and 6.

\subsection{Non-interacting limit}
For $U=0$, the impurity Green function $g(\w)$ is given by
\be
g(\omega) = \left[\omega + \mbox{i}\eta \mbox{sgn}(\omega)-\Delta(\omega)\right]^{-1}\ \ \ \ \ :\eta \rightarrow 0+
\ee
and is determined by the hybridization function
\alpheqn
\be
\Delta(\omega)=\sum_{\bi{k}}\frac{|V_{i\bi{k}}|^{2}}{\omega-\epsilon_{\bi{k}}+\mbox{i}\eta \mbox{sgn}(\omega)}\ = \Delta_{\mbox{\ssz{R}}}(\omega)-\mbox{isgn}(\omega)\Delta_{\mbox{\ssz{I}}}(\omega)
\ee
with
\be
\Delta_{\mbox{\ssz{I}}}(\w)=\pi\sum_\bi{k}|V_{i\bi{k}}|^2\delta(\w-\epsilon_{\bi{k}}).
\ee
\reseteqn
We consider a symmetric hybridization, $\Delta(\w)=-\Delta(-\w)$, and in particular a power-law form 
\be
\Delta_{\subi}(\w)=\Delta_0|\w|^r\theta(D-|\w|)
\ee
with $r \geq 0$ and bandwidth $D$ ($\theta(x)$ being the unit step function).  A pure power-law hybridization, while naturally not realistic on arbitrary scales, captures the requisite low-$\w$ behaviour in the simplest way; moreover, as familiar from the usual $r=0$ Anderson model (see e.g. [2]), one expects impurity properties to be controlled primarily by the low-$\w$ behaviour and largely independent of detailed band structure.  Note also from equation (2.4b) that to specify $\Delta_{\subi}(\w)$, the $\{V_{i\bi{k}}\}$ and host eigenvalues $\{\epsilon_{\bi{k}}\}$ do not require separate specification; but that for the particular case of constant $V_{i\bi{k}} = V$ (considered in section 6.2), $\Delta_{\subi}(\w)$ and the host spectrum $\rho_{\mbox{\ssz{host}}}(\w)$ are simply related:
\be
\Delta_{\subi}(\w)=\pi V^2 \rho_{\mbox{\ssz{host}}}(\w).
\ee

The real part of the hybridization, $\Delta_{\subr}(\w)$, follows from a Hilbert transform, viz
\be
F_{\mbox{\ssz{R}}}(\omega)=\int_{-\infty}^{\infty}\frac{\mbox{d}\omega_{1}}{\pi}\ F_{\mbox{\ssz{I}}}(\omega_{1})\ \mbox{P}\left(\frac{1}{\omega-\omega_{1}}\right)
\ee  
with $F=\Delta$; and hence
\be
\Delta_{\mbox{\ssz{R}}}(\omega)= \mbox{sgn}(\omega)\Delta_{0}|\omega|^r\frac{2}{\pi}\int_{0}^{D/|\omega|}\mbox{d} y\  \frac{y^r}{\left(1-y^2\right)}
\ee
where a principal value is henceforth understood.  We shall require explicitly only the low-$\w$ behaviour of $\Delta_{\subr}(\w)$, given from equation (2.8) (for any $r \geq 0$) by
\be
\fl\ \ \ \Delta_{\subr}(\w)= -\mbox{sgn}(\omega)\Delta_{0}\left\{\mbox{tan}\left(\mbox{$\frac{\pi}{2}r$}\right)|\omega|^r+\frac{2D^r}{\pi(r-1)}\frac{|\omega|}{D}+\Or\left[\left(\frac{|\omega|}{D}\right)^3\right]\right\}.
\ee
Notice from equation (2.8) that the wide-band limit $D \rightarrow \infty$, as commonly employed for the normal Anderson model $r=0$ (see e.g. [2]), can be taken for $r<1$; and for this case, $\Delta_{\subr}(\w)=-\mbox{sgn}(\w)\Delta_0\mbox{tan}(\frac{\pi}{2}r)|\w|^r\ \forall\  \w$.

\tolerance=20
Given the hybridization function, the non-interacting $g(\w)=\mbox{Re}g(\w)-\mbox{i}\pi\mbox{sgn}(\w)d_0(\w)$ follows directly from equation (2.3).  All relevant impurity properties are determined by the spectral density $d_0(\w)$, including `excess' thermodynamic functions induced by addition of the impurity, and local properties such as the impurity susceptibility $\chi_{ii}^0(T)=-g\mu_{\mbox{\ssz{B}}}(\partial\bra \hat{S}_{iz}\ket/\partial h)\mid_{h=0}$ (with $h$ a magnetic field acting solely on the impurity).  Details are given in [29]; here we summarize results relevant to the present work.

The key feature of the non-interacting limit is that LM states occur exclusively for $r>1$ and SC states for $r<1$; with clear signatures of the respective phases apparent in the single particle spectrum $d_0(\w)$.  For $r>1$ (LM) $d_0(\w)$ is given for $|\w|<D$ by 
\alpheqn
\be
d_0(\w)=q\delta(\w)+d_0^{\band}(\w)\ \ \ \ \ : r>1
\ee
and contains both a discrete state at $\w = 0$, with poleweight $q = \left[1-(\partial \Delta_{\subr}/\partial \w)_{\w = 0}\right]^{-1}$ given from equation (2.9) by
\be
q^{-1} = 1 + \frac{2\Delta_0 D^{r-1}}{\pi(r-1)}
\ee
\reseteqn
and a continuum (or `band') piece $d_0^{\band}(\w)\sim |\w|^{r-2}$ as $\w \rightarrow 0$.  The pole contribution to $d_0(\w)$ is the characteristic spectral signature of the $U=0$ LM state.  It produces [29] for example a local susceptibility $\chi_{ii}^0(T)=(q^2/2)\chi_{\mbox{\ssz{Curie}}}(T)$ as $T \rightarrow 0$, i.e.\ $\lim_{T\rightarrow 0}T\chi_{ii}^0(T)= q^2(g\mu_{\mbox{\ssz{B}}})^2/8$: the impurity spin remains unquenched, symptomatic of a LM state.  For $r<1$ (SC) by contrast, there is no $\w = 0$ pole contribution and $d_0(\w)\equiv d_0^{\band}(\w)$ is given as $\w \rightarrow 0$ by
\be
d_0^{\band}(\w)=\frac{|\w|^{-r}}{\pi\Delta_0 \left[1+\tan^2(\mbox{$\frac{\pi}{2}r$})\right]} + \mbox{O}(|\w|^{1-2r}) \ \ \ \ \ : r<1
\ee
for any $r \geq 0$, with a characteristic $|\omega|^{-r}$ divergence for $0<r<1$. In consequence [29], $\lim_{T\rightarrow 0} T\chi_{ii}^0(T) = 0$: the spin is quenched as occurs for the normal $r=0$ Anderson model, one reason why the SC state may be regarded [21,29] as a natural generalization of conventional Fermi liquid physics.

\subsection{SC state: spectral pinning}
For $U>0$ the impurity Green function, $G(\w)=X(\w)-\mbox{i}\pi\mbox{sgn}(\w)D(\w)$ ($=-G(-\w)$ by particle-hole symmetry), may be expressed as
\be
G(\omega) = \left[\omega + \mbox{i}\eta \mbox{sgn}(\omega)-\Delta(\omega)-\Sigma(\w)\right]^{-1}
\ee
where
\be
\Sigma(\w) = \Sigma_{\subr}(\w)-\mbox{i}\mbox{sgn}(\w)\Sigma_{\subi}(\w)
\ee
is an interaction self-energy whose real/imaginary parts are related by the Hilbert transform equation (2.7) with $F=\Sigma$. (Note that $\Sigma(\w)$ is defined to exclude the trivial Hartree contribution: from particle-hole symmetry the Fermi level remains fixed at $\w =0$ for all $U \geq 0$ (and any $r$), whence the impurity charge $n_i = \sum_\sigma \bra\hat{n}_{i\sigma}\ket = 1\ \forall \ U$ and thus the Hartree contribution of $(U/2)n_i=U/2$ trivially cancels $\epsilon_i = -U/2$.)

In [29] we have established conditions upon $\Sigma(\w)$ for a SC state to arise for $U>0$.  These are, very simply, that $\Sigma_{\subi}(\w)$ (and hence $\Sigma_{\subr}(\w)$) should decay to zero as $\w \rightarrow 0$ more rapidly than $|\w|^r$ (with $r<1$), i.e.
\be
\Sigma_{\subi}(\w)\stackrel{\w\rightarrow 0}{\sim}\alpha|\w|^\lambda\ \ \ \ \ :\lambda>r\ee
with $\lambda > r$.  In consequence, as follows directly from equation (2.12), the low-frequency behaviour of $D(\w)$ is that of $d_0(\w)$; hence from equation (2.11), $D(\w)\stackrel{\w\rightarrow 0}{\sim}|\w|^{-r}$ --- which is  indeed the spectral signature of the SC state found in finite-$U$ NRG studies [28].  Further, using
\be
\lim_{\w \rightarrow 0}|\w|^rD(\w)=\lim_{\w \rightarrow 0}|\w|^rd_0(\w)
\ee
and defining the modified spectral function
\alpheqn
\be
A(\w)=|\w|^rD(\w)
\ee
equation (2.11) yields the pinning condition
\be
\pi \Delta_0 \left[1+\mbox{tan}^2\mbox{$(\frac{\pi}{2}r)$}\right]A(\w=0)=1
\ee
\reseteqn
for all $U$ and $r$ where a SC state obtains.  This result will prove central to our analysis in the following sections.  It encompasses as a special case the well known result for the normal $r=0$ Anderson model (see e.g.[2]): that $\pi\Delta_0 D(\w =0)=1$ --- the impurity spectrum is pinned at the Fermi level $\w =0$ for any $U$ where a normal Fermi liquid state obtains (in that case all $U \geq 0$).  Equation (2.16) generalizes the pinning condition to arbitrary $r$ for a SC state, and reflects the fact that interactions have no influence in renormalizing the asymptotic behaviour of $D(\w)$ as $\w \rightarrow 0$, again consistent with the view [21,29] that the SC state constitutes a natural generalization of Fermi liquid behaviour.
\seceq
\section{Two-self-energy description}
Equation (2.12) merely defines a single self-energy $\Sigma(\w)$, via a Dyson equation
\be
G(\w) = g(\w)+g(\w)\Sigma(\w)G(\w)
\ee
that does not by itself enable a calculation of the impurity Green function.  And while at first sight it may invite a perturbative treatment in $U$ about the non-interacting limit (where $G(\w)\equiv g(\w)$), there are two reasons to be wary of such an approach.  First, the general applicability of such is not obvious in the soft-gap problem: indeed, as discussed in [29], there is evidence to suggest that for $r \in [\frac{1}{2},1]$, perturbation theory in $U$ is inapplicable as $U \rightarrow 0$ (a point to which we return again in section 7.1).  Second, and more generally, even if a perturbative approach in $U$ is possible for sufficiently low $U$ --- as it is known to be [29] for $0 \leq r <\frac{1}{2}$ and $r>1$ --- straight perturbation theory in the interaction strength naturally cannot capture the transition between SC and LM states.  For this, one requires an inherently non-perturbative approach that is capable of describing both the doubly degenerate LM state and the SC (or generalized Fermi liquid) state, and hence the quantum phase transition between them.

To this end, and noting that direct calculation of a single self-energy is not sacrosanct, we follow recent work [30] on the normal $r=0$ Anderson model and adopt a two-self-energy description, with the impurity Green function expressed formally as
\be
G(\w)=\mbox{$\frac{1}{2}$}\left[G_\up(\w)+G_\down(\w)\right]
\ee
where
\be
G_\sigma(\w)=\left[\w+\mbox{i}\eta\mbox{sgn}(\w)-\Delta(\w)-\tilde{\Sigma}_\sigma(\w)\right]^{-1}
\ee
with interaction self-energies $\tilde{\Sigma}_\sigma(\w)$ (and $\sigma = \up/\down$ or $+/-$).  Such a description is self-evidently natural to describe a doubly degenerate LM state and, more generally, is appropriate if one seeks to construct a non-perturbative theory starting from either the atomic limit ($V_{i\bi{k}}=0$, where $\Sigma_\sigma(\w)=-\frac{1}{2}\sigma U$) or an unrestricted Hartree Fock (UHF) mean-field approach.  It is the latter strategy that we adopt and, although the deficiences of a simple UHF approach by {\it itself} are significant, use of it as a starting point for a genuine many-body treatment will be shown to yield a rather successful description of the soft-gap problem, as well as the normal $r=0$ Anderson model considered hitherto [30].

For the symmetric case under consideration, particle-hole symmetry implies
\be
G_\up(\w)=-G_\down(-\w)
\ee
(and thus $D_\up(\w)=D_\down(-\w)$ for the corresponding spectral densities $D_\sigma(\w)=-\pi^{-1}\mbox{sgn}(\w)\mbox{Im}G_\sigma(\w)$); from which, since $\Delta(\w)=-\Delta(-\w)$, equation (3.3) implies
\be
\tilde{\Sigma}_\up(\w)=-\tilde{\Sigma}_\down(-\w).
\ee
In consequence, from equations (3.2) and (2.12), the impurity Green function and single self-energy satisfy the familiar conditions
\be
G(\w)=-G(-\w), \ \ \ \ \Sigma(\w)=-\Sigma(-\w).
\ee
Equations (3.4-6) merely express a basic symmetry, which must of course be satisfied by any approximate theory; and equation (3.5) in particular shows that it is sufficient to consider only one of the $\tilde{\Sigma}_\sigma(\w)$, say $\tilde{\Sigma}_\up(\w)$.  Once a theory for $\tilde{\Sigma}_\up(\w)$ has been developed, direct comparison of equations (3.2,3) with equation (2.12) permits, if desired, $\Sigma(\w)$ to be determined, via
\be
\Sigma(\w)=\frac{\frac{1}{2}\left[\tilde{\Sigma}_\up(\w)-\tilde{\Sigma}_\up(-\w)+2g(\w)\tilde{\Sigma}_\up(\w)\tilde{\Sigma}_\up(-\w)\right]}{1-\frac{1}{2}g(\w)\left[\tilde{\Sigma}_\up(\w)-\tilde{\Sigma}_\up(-\w)\right]}
\ee
where $g(\w)$ is the non-interacting Green function, equation (2.3).

The $\tilde{\Sigma}_\sigma(\w)$ are obviously not calculable exactly, but diagramatic perturbation theory based upon a UHF mean-field state can be employed to develop suitable (and indeed asymptotically exact) approximations as detailed in section 5.  To this end it is helpful to separate the full interaction self-energies as
\be
\tilde{\Sigma}_\sigma(\w)=-\frac{\sigma}{2}U|\mu|+\Sigma_\sigma(\w)
\ee
where $\pm \frac{1}{2}U|\mu|$ is the purely static Fock contribution which alone survives at UHF level (with $|\mu|$ the local moment magnitude); and where the $\Sigma_\sigma(\w)$ --- to which the symmetry equation (3.5) also applies --- contains the dynamics that, at low frequencies in particular, are naturally central to the problem.

Before proceeding we note that conditions upon $\tilde{\Sigma}_\sigma(\w)$ for a SC state to arise for $U>0$ are readily established, independently of any specific approximation.  As in section 2.2 we consider explicitly $r<1$ (since a SC state must be perturbatively continuable from the non-interacting limit and for $r>1$ the $U=0$ ground state is a LM one [29]).  The requisite conditions upon $\tilde{\Sigma}_\sigma(\w)$ are identical to those of section 2.2 for $\Sigma(\w)$: the real and imaginary parts of $\tilde{\Sigma}_\sigma(\w)$ must decay to zero as $\w \rightarrow 0$ more rapidly than $\Delta_{\subi/\subr}\sim |\w|^r$.  From equations (3.2,3) the $\w \rightarrow 0$ behaviour of $G_\up(\w),\ G_\down(\w)\ \mbox{and}\ G(\w)$ then coincide, and reduce to that of the non-interacting limit; equation (2.15) is thus satisfied, and in consequence the generalized spectral pinning condition characteristic of the SC state, equation (2.16b), follows.

Finally, notice that a necessary condition for a SC state to arise is thus $\tilde{\Sigma}_\sigma(\w=0)=0$ (which from equation (3.5) is independent of spin, $\sigma$); or equivalently, from equation (3.8):
\be
\Sigma_\up(\w=0)=\frac{1}{2}U|\mu|.
\ee
The practical importance of this condition will become apparent in section 5, for while necessary but not {\it a priori} sufficient for a SC state, its imposition as a self-consistency condition underlies our analysis of the SC phase, in direct parallel to our previous work on the normal $r=0$ Anderson model [30].
\seceq
\section{Mean field}
We start from the simplest non-trivial mean-field approximation, viz UHF as considered in Anderson's original paper [1].  This has two essential characteristics: that the notion of an impurity local moment ($\mu$) is introduced explicitly from the outset; and that it is determined self-consistently, via $\mu =\bra \hat{n}_{i\up}-\hat{n}_{i\down}\ket_0$ (with $\bra ...\ket_0$ an average over the mean-field ground state).  There are three reasons for first considering this superficially simple one-body approximation.  First, the UHF Green functions form the bare propagators for the dynamical many-body approach developed in sections 5ff.  Second, a physical understanding of its limitations, often alluded to but rarely exposed, underpins what is required to go successfully beyond it.  Finally, we show that even UHF by itself has virtues, producing a number of non-trivial predictions that are in qualitative accord with sophisticated approaches; notably that {\it only} LM states arise for $r>\frac{1}{2}$, in agreement with detailed NRG calculations [21,28]. 

The essence of UHF is that the self-energies $\tilde{\Sigma}_\sigma(\w)$ (equation (3.8)) are purely static: only the Fock term is retained, and $\tilde{\Sigma}^0_\sigma=-\frac{1}{2}\sigma U|\mu|$.  The impurity Green function at UHF level, $G_0(\w)=\mbox{Re}G_0(\w)-\mbox{i}\pi \mbox{sgn}(\w)D_0(\w)$, is thus
\be
G_0(\w)=\mbox{$\frac{1}{2}$}\left[{\cal G}_\up(\w)+{\cal G}_\down(\w)\right] 
\ee
where
\be
{\cal G}_\sigma(\w)=\left[\w+\mbox{i}\eta \mbox{sgn}(\w)+\frac{\sigma}{2}U|\mu|-\Delta(\w)\right]^{-1}
\ee
with corresponding spectral densities $D^0_\sigma(\w)=-\pi^{-1} \mbox{sgn}(\w)\mbox{Im}{\cal G}_\sigma(\w)$ given by
\be
D^0_\sigma(\w)=\frac{\left[\eta+\Delta_{\subi}(\w)\right]\pi^{-1}}{\left[\w+\frac{\sigma}{2}U|\mu|-\Delta_{\subr}(\w)\right]^2+\left[\eta+\Delta_{\subi}(\w)\right]^2}.
\ee
Quite generally there are three energy scales in the problem, viz $\delr, U$ and $D$; we choose to rescale in terms of $\delr$, defining for later purposes a reduced interaction strength and bandwidth respectively by
\be
\tilde{U}=\frac{U}{\delr}\ \ \ \ \tilde{D}=\frac{D}{\delr}
\ee
as well as a dimensionless frequency, $\tilde{\w}=\w / \delr$.

\subsection{Mean-field phase boundary}
The local moment $|\mu|$ is of course found self-consistently, as described below.  If $|\mu|=0$ thereby arises, then from equations (4.2,3) the UHF propagators and spectra reduce to those of the non-interacting limit summarized in section 2.1: $D^0_\sigma(\w)\equiv d_0(\w)$.  If $|\mu|\neq 0$ by contrast, equation (4.3) (with equations (2.5,9) for $\Delta_{\subi / \subr}$) show the low-frequency behaviour of $D^0_\sigma(\w)$ to be
\be
D^0_\sigma(\w)\stackrel{\w \rightarrow 0}{\sim}\frac{\Delta_0}{\pi\left(\frac{1}{2}U|\mu|\right)^2}|\w|^r
\ee
independent of spin $\sigma$.  Equation (4.5) thus gives the low-$\w$ behaviour of the full UHF spectrum $D_0(\w)=\frac{1}{2}\left[D^0_\up(\w)+D^0_\down(\w)\right]$; and we note immediately that $D(\w) \sim |\w|^r$ is in fact the spectral hallmark of the LM regime obtained from finite-$U$ NRG calculations [28].  We also add in passing that for finite bandwidth $D$, and regardless of $U$, there are always discrete (pole) contributions to the $D^0_\sigma(\w)$ outside the band, $|\w|>D$.  These are included in all specific calculations (section 6 ff), but are of little importance to the problem and are not discussed explicitly in what follows.

At pure UHF level the local moment is determined self-consistently from
\be
|\mu|=\int_{-\infty}^0\mbox{d}\w\left[D^0_\up(\w)-D^0_\down(\w)\right].
\ee
Noting from equation (4.3) that the $D^0_\sigma(\w)$ depend on $U$ and $|\mu|$ solely via the combination
\be
x = \frac{1}{2}U|\mu|
\ee
equation (4.6) is thus of form
\be
(|\mu| = )\ \frac{2x}{U}=f(x).
\ee
The UHF phase boundary $U_{\cc} \equiv U^0_{\cc}(r)$ is now readily ascertained from the $x \rightarrow 0$ behaviour of $f(x)$, noting that $x \rightarrow 0$ may correspond either to (i) $|\mu| \rightarrow 0$ at some finite critical $U_{\cc}$ as one might naively anticipate; or (ii) $U_{\cc} =0$ and $|\mu|$ either vanishing or remaining finite as $U \rightarrow 0$.  With
\alpheqn
\be
f(x)\stackrel{x \rightarrow 0}{\sim}x^m
\ee
and hence from equation (4.8)
\be
\frac{1}{U}\stackrel{x \rightarrow 0}{\sim}x^{m-1}
\ee
\reseteqn
the above possibilities are distinguished by different values of the exponent $m$.  If $m=1$ then $U_{\cc}$ is finite and given by
\alpheqn
\be
\frac{2}{U_{\cc}}=\left(\frac{\partial f(x)}{\partial x}\right)_{x=0}\ \ \ \ \ :m = 1.
\ee
If by contrast $0<m<1$, then equation (4.9b) shows $U_{\cc} =0$ and
\be
|\mu|\stackrel{U \rightarrow 0}{\sim}U^{\frac{m}{1-m}} \ \ \ \ \ : 0<m<1
\ee
vanishes as $U \rightarrow 0$.  Finally, if $m=0$, then $U_{\cc} =0$ but
\be
|\mu|\stackrel{U \rightarrow 0}{\sim}\mbox{finite} \ \ \ \ \ : m=0
\ee
\reseteqn
tends to a finite limit as $U \rightarrow 0+$.

All three posibilities are realized in practice, as now summarized (details are given in the Appendix).  For $r<\frac{1}{2}, m=1$ results: there is a finite critical $U^0_{\cc}(r)$ separating LM states ($U>U^0_{\cc}(r), |\mu|>0$) from SC states ($U<U^0_{\cc}(r), |\mu|=0$).  For $r \in (\frac{1}{2},1)$ by contrast, $0<m=(1-r)/r<1$; while $m=0$ for $r>1$.  It follows directly that for $r<\frac{1}{2}$ both SC and LM states may arise, while for $r>\frac{1}{2}$ exclusively LM states occur for all $U>0$.  This is as found in finite-$U$ NRG calculations [21,28].  It is moreover specific to the particle-hole symmetric case under consideration: for the asymmetric case, analysis of the UHF equations yields a finite $U^0_{\cc}(r)$ even for $r>\frac{1}{2}$; this is again in qualitative agreement with NRG results [21].

The resultant mean-field phase boundary is shown explicitly in Figure 1 for the wide-band limit $D = \infty$ (which depends solely on the ratio $\tilde{U}=U/\delr$); for later comparison to NRG results, the figure shows the critical $\Delta_0U^r/U (=\tilde{U}^{r-1}$), versus $r$.  For $r<\frac{1}{2}$, the critical $U^0_{\cc}(r)$ is finite and given from equation (4.10a) using equations (4.8,6) and (4.3) by
\be
\frac{2}{U^0_{\cc}(r)}=-\frac{4}{\pi}\int_{-\infty}^0 \mbox{d}\w\frac{\Delta_{\subi}(\w)\left[\w-\Delta_{\subr}(\w)\right]}{\left(\left[\w-\Delta_{\subr}(\w)\right]^2+\Delta^2_{\subi}(\w)\right)^2}\ .
\ee
\begin{figure}
\begin{center}
\epsfig{file =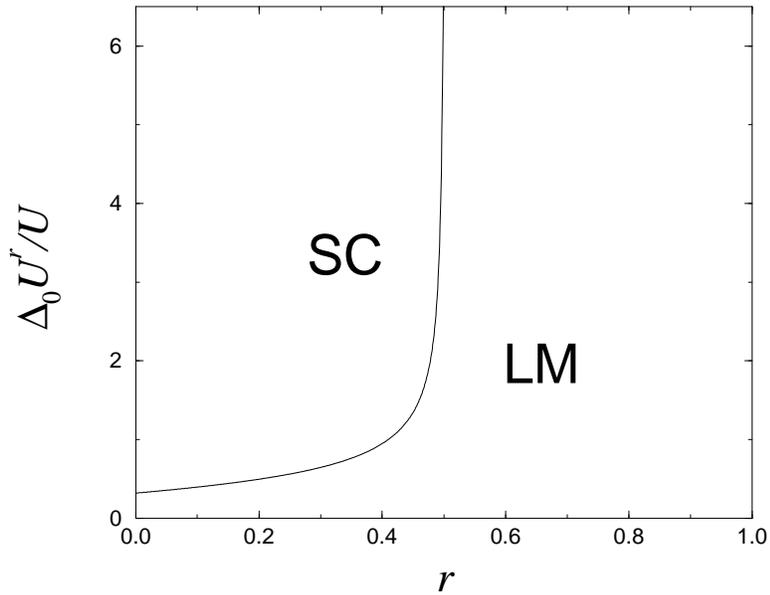,width=10cm} 
\caption{Mean-field phase boundary $(\Delta_0U^r/U)_{\cc}$ verus $r$ (for wide-band limit).  For $r>\frac{1}{2}$, solely LM states occur for all $U>0$.}
\end {center}
\end{figure}
Since $\Delta_{\subi / \subr}\sim |\w|^r$ as $\w \rightarrow 0$, the low-$\w$ behaviour of the integrand is $\sim |\w|^{-2r}$; the integral thus converges for $r<\frac{1}{2}$, and $U^0_{\cc}(r)\sim (1-2r)$ as $r \rightarrow \frac{1}{2}-$, producing the square root divergence in the phase boundary evident in Figure 1: $\Delta_0U^r/U \sim (1-2r)^{-1/2}$.

For $r>\frac{1}{2}$ by contrast, even the simple mean-field analysis predicts solely a LM phase for any $U>0$.  Recall however that for the non-interacting limit $U=0$, the ground state is LM for $r>1$ but SC for $r<1$ [29] (see section 2.1).  Hence for $\frac{1}{2}<r<1$, there is a critical line $U_{\cc}(r) = 0$ `separating' SC and LM states.  This underlies why low-order perturbation theory in $U$ about the non-interacting limit is inapplicable for $r \in [\frac{1}{2},1]$ [29] (see also section 7.1); and the subtlety of $U \rightarrow 0$ is already evident at mean-field level by the fact that the UHF $|\mu|$ vanishes in a non-analytic fashion, viz
\be
|\mu|\stackrel{U \rightarrow 0}{\sim}U^{\frac{1-r}{2r-1}} \ \ \ \ \ :r \in \mbox{$(\frac{1}{2},1)$}.
\ee
For $r>1$ by comparison, the ground state is LM for all $U\geq 0$; and $|\mu|$ remains finite as $U \rightarrow 0$ (equation (4.10c)).  In fact, as shown in the Appendix, the $U=0+$ limit of $|\mu|$ is
\be
|\mu| = q \ \ \ \ \ :U=0+,\ r>1
\ee
where $q$ (equation (2.10b)) is precisely the weight of the $\w=0$ pole in the non-interacting single-particle spectrum $d_0(\w)$.

Finally, we clarify the familiar remark that UHF is a {\it static} mean-field approximation.  In one sense this is trivial: the self-energies $\tilde{\Sigma}_\sigma(\w)$ (equation (3.8)) are $\w$-independent at UHF level, being given by $\tilde{\Sigma}^0_\sigma=-\frac{1}{2}\sigma U|\mu|$.  But the corresponding {\it single} self-energy $\Sigma(\w)$, defined conventionally via $G(\w)=[\w + \mbox{i}\eta\mbox{sgn}(\w)-\Delta(\w)-\Sigma(\w)]^{-1}$, is given from equation (3.7) by
\be
\Sigma^{\mbox{HF}}(\w)=g(\w)(\mbox{$\frac{1}{2}$}U|\mu|)^2
\ee
with $g(\w)$ the non-interacting Green function, equation (2.3).  Thus, even at UHF level, $\Sigma(\w)$ is $\w$-{\it dependent}.  For $r>1$ in particular, the leading low-$\w$ behaviour of $g(\w)$ is given from equations (2.3,10) by $g(\w)\sim q/(\w+\mbox{i}\eta \mbox{sgn}(\w))$.  Hence, from equation (4.13), the leading low-$\w$ behaviour of $\Sigma^{\mbox{HF}}(\w)$ as $U \rightarrow 0$ is given by
\be
\Sigma^{\mbox{HF}}(\w)\stackrel{\w \rightarrow 0}{\stackrel{U \rightarrow 0}{\sim}}\frac{U^2q^3}{4}\frac{1}{\w+\mbox{i}\eta \mbox{sgn}(\w)} \ \ \ \ \  :r>1.
\ee
And simple though it is, this result is not trivial: it recovers exactly the leading low-$\w$ behaviour of $\Sigma(\w)$ obtained [29] from second order perturbation theory in $U$ about the non-interacting limit (which is applicable for $r>1$ [29]).

\subsection{Deficiences}
Its virtues notwithstanding, the limitations of UHF by itself are of course severe.  If the self-consistent mean-field local moment $|\mu|=0$, UHF reduces trivially to the non-interacting limit; there is thus no hint of the low-energy Kondo scale symptomatic of the $r \geq 0$ SC phase and evident in the generalized Kondo resonance appearing in the modified spectral function $A(\w)=|\w|^rD(\w)$ [29].  But the acute deficiences of simple mean-field are already evident in the UHF phase boundary of Figure 1.  From finite-$U$ NRG studies [21,28] it is known that the critical $\Delta_0U^r/U$ vanishes linearly in $r$ as $r \rightarrow 0$ i.e. $\Delta_{0,\cc}\sim r$ or $U_{\cc} \sim 1/r$.  For $r=0$ this recovers the well known fact that the normal Anderson model is a Fermi liquid for any non-zero hybridization strength $\Delta_0$ (or finite $U$), with a LM phase confined exclusively to the atomic limit, $\Delta_0 = 0$.  This is not however captured by UHF, which instead produces a critical $\Delta_0/U = 1/\pi$ for $r=0$ (see Figure 1) and thus a spurious transition between the SC (or Fermi liquid) and LM phases at a finite coupling strength.

More generally, the $r \rightarrow 0$ behaviour of the NRG phase boundary for the Anderson model [21,28] is indicative of that for the corresponding Kondo model as considered originally by Withoff and Fradkin [22]; for since $\Delta_0/U_{\cc} \propto r$ as $r \rightarrow 0$, the Anderson model can here be mapped onto a Kondo model [21,28] via the usual Schrieffer-Wolf transformation [32], with an exchange coupling $J \propto V^2_{i\bi{k}}/U$.  Thus, as found originally via poor man's scaling for the Kondo model itself [22], there exists an infrared unstable fixed point at $J_{\cc} \propto r$ such that for $J>J_{\cc}$ ($J<J_{\cc}$) the ground state is SC(LM).  Clearly, however, no vestige of this Kondo physics is captured at UHF level.

The manifest deficiences of the simple mean-field approximation naturally stem from its static character, and in particular from the complete omission of dynamical spin-flip scattering processes illustrated schematically in figure 2.  Upon addition of, say, a $\down$-spin electron to an $\up$-spin occupied impurity, two subsequent processes may occur.  (a) The added $\down$-spin may hop off the site, leaving behind the original $\up$-spin.  This is essentially elastic scattering; it is well captured by UHF alone.  (b) However the $\up$-spin electron originally present may also hop off the site, leaving behind a spin-flip on the impurity, the energy cost for the spin-flip being on the order of the Kondo scale.  This process, involving correlated electron motion and dynamical coupling of single-particle excitations to low-energy spin fluctuations, is entirely absent at UHF level.  Inclusion of such, to which we now turn, is however essential to circumvent {\it all} the limitations of the static mean-field approach outlined above, and in particular to recover the correct physics of the Kondo (or spin-fluctuation) regime.  On the other hand one should not abandon UHF entirely, but rather use it as a starting point for a dynamical many-body approach; use of it in this fashion is, as we shall show, necessary to ensure a successful description of the problem for all $r \geq 0$ and from weak to strong coupling interaction strengths, $U$.
\begin{figure}
\begin{center}
\epsfig{file =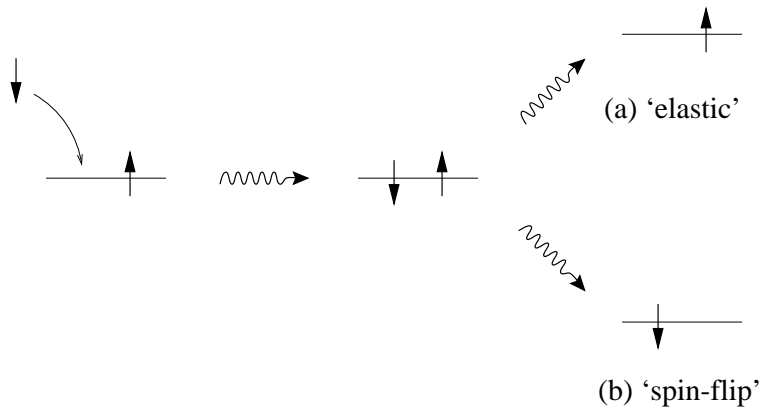,width=10cm} 
\caption{Schematic of scattering processes as discussed in text.}
\end {center}
\end{figure}

\seceq
\section{Dynamical self-energies}
The interaction self-energies $\tilde{\Sigma}_\sigma(\w)$ consist, as in equation (3.8), of a static Fock contribution (alone retained at mean-field level) plus the dynamical contribution $\Sigma_\sigma(\w)$ on which we now focus.  The most important class of diagrams contributing to the $\Sigma_\sigma(\w)$, and that we retain in practice, is shown in Figure 3a; mean-field impurity propagators (given by equation (4.2)) are denoted by solid lines, and the impurity interaction $U$ by a wavy line.  The physical content of Figure 3a is clear: having, say, added a $\sigma$-spin electron to a -$\sigma$-spin occupied impurity, the latter hops off the impurity, generating in consequence an on-site spin-flip, before returning again at a later time; and where all ladder interactions of the resultant particle-hole pair --- reflecting the created spin-flip --- are included.  This class of diagrams thus captures the dynamical spin-flip scattering mentioned above (Figure 2b), and known e.g. from poor man's scaling [33] to be essential in describing the Kondo limit of the normal $r=0$ Anderson model.  This is further evident from the equivalent recasting of $\Sigma_\sigma(\w)$ shown in Figure 3b, which translates to 
\alpheqn
\be
\Sigma_{\up}(\w)=U^2 \int_{-\infty}^{\infty}\frac{\mbox{d}\w_1}{2\pi \mbox{i}}\ {\cal G}_\down(\w-\w_1)\Pi^{-+}(\w_1)
\ee
\be
\Sigma_{\down}(\w)=U^2 \int_{-\infty}^{\infty}\frac{\mbox{d}\w_1}{2\pi \mbox{i}}\ {\cal G}_\up(\w-\w_1)\Pi^{+-}(\w_1).
\ee
\reseteqn
\begin{figure}
\begin{center}
\epsfig{file =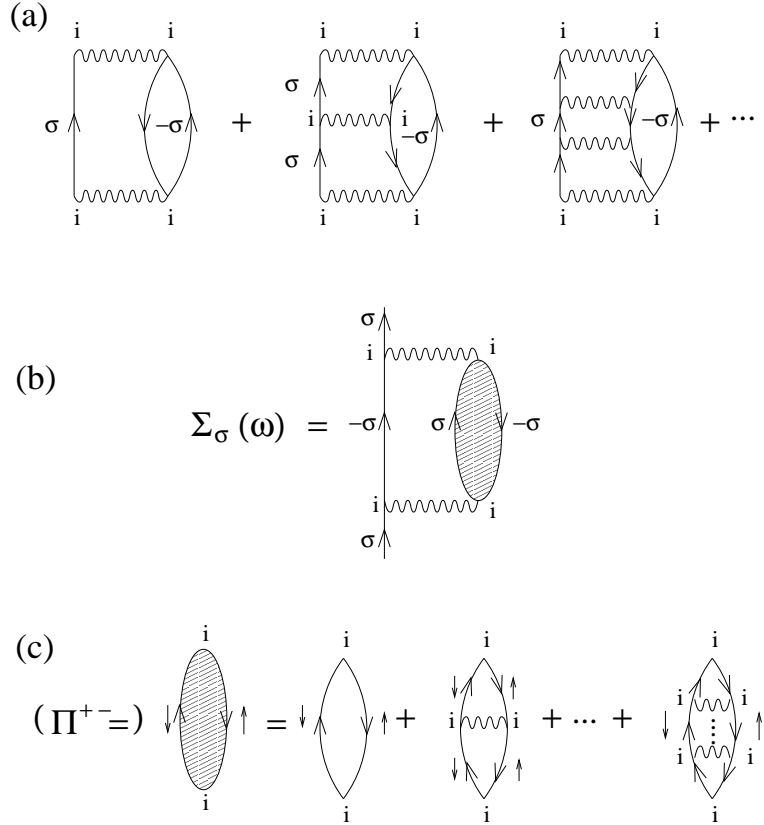,width=10cm} 
\caption{(a) Class of diagrams for $\Sigma_\sigma(\w)$ retained in present work.  Mean-field impurity propagators are denoted by solid lines, the on-site impurity $U$ by a wavy line.  (b) Equivalent recasting, including ingoing/outgoing propagators, to illustrate spin-flip scattering involved. (c) Particle-hole ladder sum in transverse spin channel; for $\Pi^{-+}$, spins are reversed.}
\end {center}
\end{figure}
These embody dynamical coupling of single-particle excitations to low-energy spin fluctuations, since the transverse spin polarization propagators $\Pi^{+-}/\Pi^{-+}$ --- given as in Figure 3c by the ladder sum of repeated particle-hole interactions in the transverse spin channel --- contain as will be shown the low-energy spin flip scales that are the essence of {\it both} the SC and LM phases.

The transverse spin polarization propagators are given in turn by an RPA form
\be
\Pi(\w)=\frac{\ ^0\!\Pi(\omega)}{1-U\ ^0\!\Pi(\omega)}
\ee
with $^0\!\Pi$ the bare polarization bubble (first diagram in Figure 3c).  $^0\!\Pi^{+-}(\omega)$ is given explicitly by
\be
\ ^0\!\Pi^{+-}(\omega)=\mbox{i}\int_{-\infty}^\infty\frac{\mbox{d}\w_1}{2\pi}\ {\cal G}_\down(\w_1){\cal G}_\up(\w_1-\w)
\ee
while $^0\!\Pi^{-+}(\omega)$ follows by interchanging the spin labels, $\up \leftrightarrow \down$; and a simple change of variables in equation (5.3) gives
\be
\ ^0\!\Pi^{+-}(\omega)=\ ^0\!\Pi^{-+}(-\omega)
\ee
 --- which naturally applies also to the full $\Pi^{+-}(\w)/\Pi^{-+}(\w)$.  Only one propagator, say $\ ^0\!\Pi^{+-}(\omega)$, need thus be considered explicitly; the other follows from it.  Further, since the real/imaginary parts of $\ ^0\!\Pi^{+-}(\omega)$ are related by the Hilbert transform
\be
\ ^0\!\Pi^{+-}(\omega)=\int_{-\infty}^{\infty}\frac{\mbox{d}\w_1}{\pi}\ \frac{\mbox{Im}\ ^0\!\Pi^{+-}(\omega_1)\mbox{sgn}(\w_1)}{\w_1-\w-\mbox{i}\eta\mbox{sgn}(\w)}
\ee
we may focus on $\mbox{Im}\ ^0\!\Pi^{+-}(\omega)$.  From equation (5.3), separating ${\cal G}_\sigma(\w)={\cal G}^+_\sigma(\w)+{\cal G}^-_\sigma(\w)$ into retarded/advanced components and using the Hilbert transform
\be
{\cal G}^\pm_\sigma(\w)=\int_{-\infty}^{\infty}\mbox{d}\w_1\ \frac{D^0_\sigma(\w_1)\theta(\pm \w_1)}{\w -\w_1 \pm \mbox{i}\eta}
\ee
this is given by
\bea
\frac{1}{\pi}\mbox{Im}\ ^0\!\Pi^{+-}(\omega)=\theta(\w)&\int_0^{|\w|}\mbox{d}\w_1\ D^0_\down(\w_1)D^0_\up(\w_1-\w)\nonumber \\
&+\theta(-\w)\int_{-|\w|}^{0}\mbox{d}\w_1\ D^0_\down(\w_1)D^0_\up(\w_1-\w)\ \ \ \ \  \geq 0.
\eea

Finally, note that (equations (5.1)) preserve --- as they must --- the particle-hole symmetry $\Sigma_\down(\w)=-\Sigma_\up(-\w)$ (equations (3.5,8)); as follows using ${\cal G}_\up(\w)=-{\cal G}_\down(-\w)$ (equations (4.2), (3.4)) with equation (5.4) for the $\Pi$'s.  In what follows we thus focus exclusively on $\Sigma_\up(\w)$ which, using equation (5.4), may be written as
\be
\Sigma_\up(\w)=U^2\int_{-\infty}^\infty \frac{\mbox{d}\w_1}{2\pi \mbox{i}}\ \Pi^{+-}(\w_1){\cal G}_\down(\w_1+\w)
\ee
and which we now consider separately for both SC and LM phases.

\subsection{SC state}
We begin with a brief overview of our approach to the SC phase, and give further details below and in section 6ff.  As noted in section 3, equation (3.9) constitutes a necessary condition for a SC state to arise for $U>0$; we show below that it is also sufficient, and reduces to
\be
\Sigma^{\subr}_\up(\w=0)=\frac{1}{2}U|\mu|
\ee
(where $\Sigma^{\subr}_\sigma(\w)=\mbox{Re}\Sigma_\sigma(\w)$).  If this equation is satisfied then the generalized pinning condition symptomatic of a SC state, equation (2.16b), will be satisfied.  And the core of our approach to the SC state is to enforce equation (5.9) --- which refers to a single frequency, the Fermi level $\w=0$ --- as a self consistency condition.  In practice, as for the $r=0$ model considered hitherto [30], this amounts to a self-consistent determination of the local moment, $|\mu|$; for equation (5.9) is of form
\be
g(U;x)=x
\ee
where $g(U;x)\equiv \Sigma^{\subr}_\up(\w =0)$ depends explicitly on $U$, and upon $x=\frac{1}{2}U|\mu|$ (via the dependence of the mean-field propagators ${\cal G}_\sigma(\w)$ upon $x$).  With a chosen approximation for $\Sigma_\up(\w)$ --- equation (5.8) in the present work --- the modus operandi is clear:  for given $r$ and a chosen $U$, solve equations (5.9,10) for $x$ and hence $|\mu|$; if a solution is possible one has a SC state, and the $U$ above which a solution is no longer possible gives the critical $U_{\cc}(r)$ for termination of the SC phase, i.e. the phase boundary between SC and LM states (on the assumption, indeed found in practice (section 6), that solely LM states arise for $U>U_{\cc}(r)$).

An initial illustration of what results is seen in Figure 4 where, for $r=0.2$ (and the wide-band limit $D=\infty$), we show the resultant spectral density of transverse spin excitations, $\mbox{Im}\Pi^{+-}(\w)$ versus $\tilde{\w}=\w/\delr$, for $\tilde{U}=U/\delr =$ 9, 10 and 13.  The inset to Figure 4 shows for comparison the corresponding $\mbox{Im}\ ^0\!\Pi^{+-}(\omega)$ associated with the bare polarization bubble, shown explicitly for $\tilde{U}=13$ only (since those for the other $\tilde{U}$'s differ insignificantly from it).  Two principal points should be noted.  First, as one expects, $\mbox{Im}\ ^0\!\Pi^{+-}(\omega)$ consists simply of a high energy Stoner band centred on $\tilde{\w}\sim \tilde{U}|\mu|\sim 10^1$.  For $\mbox{Im}\Pi^{+-}(\w)$ by contrast, it is seen that the vast majority of the spectral weight has been transferred to a low-$\w$ resonance peaked at a characteristic spin-flip scale $\tilde{\w}_{\m}$ that is at least $3$ orders of magnitude smaller in the examples shown: this is essentially the Kondo scale characteristic of the SC state; it will be investigated in detail in the following sections.  Second, upon increasing $\tilde{U}$ in the SC phase, $\w_{\m}$ progressively diminishes, and vanishes at the critical $\tilde{U}_{\cc}(r) (\simeq 15.8)$ where the resonance in $\mbox{Im}\Pi^{+-}(\w)$ becomes an isolated pole at $\w=0$ precisely; as discussed below and in section 5.2, the latter is the characteristic signature of the doubly degenerate local moment state.

\begin{figure}
\begin{center}
\epsfig{file =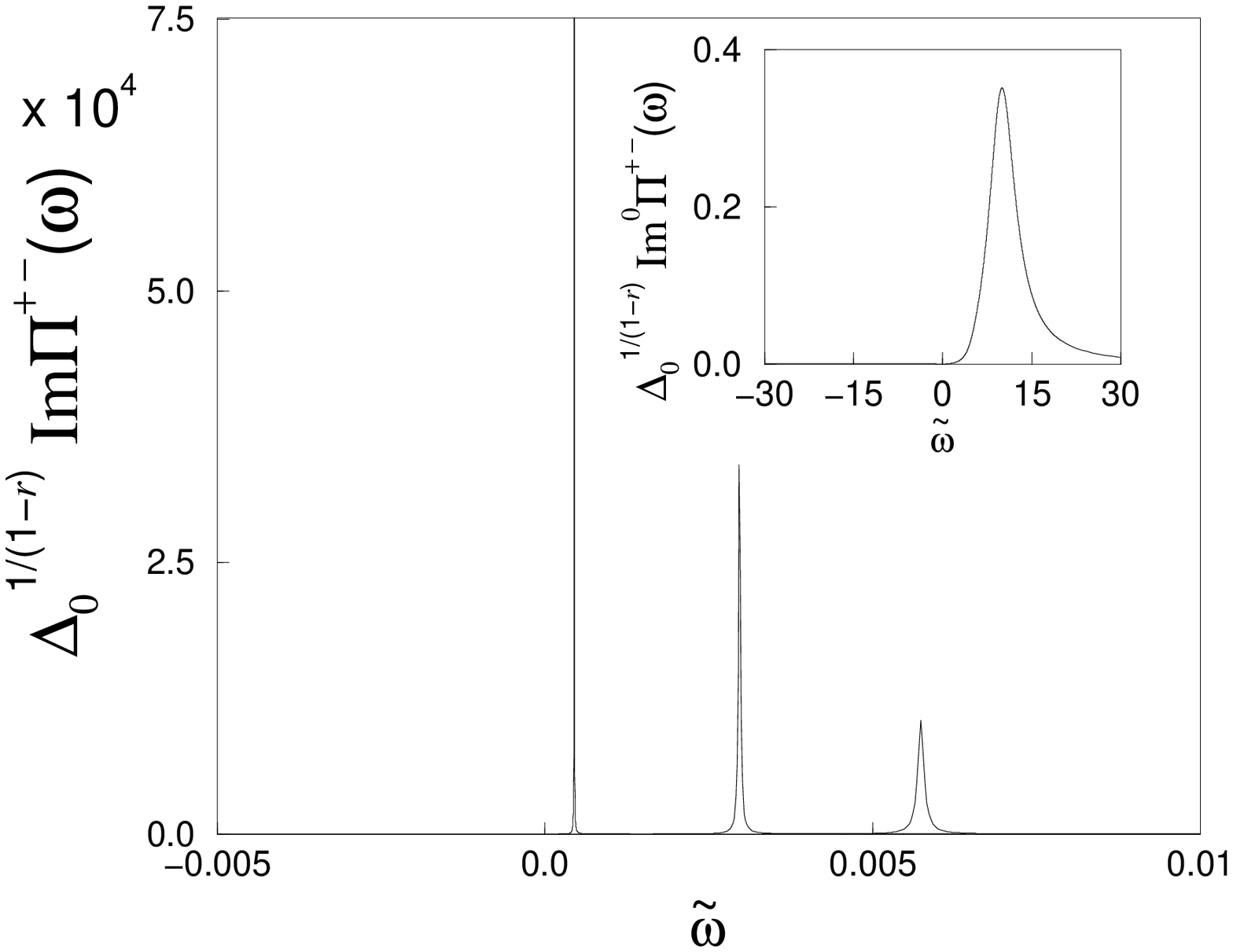,width=10cm} 
\caption{Im$\Pi^{+-}(\w)$ versus $\tilde{\w}=\w/\delr$ for $r = 0.2$ and $\tilde{U}=U/\delr = 9, 10$ and $13$ (right to left); there are no further spectral features outside range shown.  The spin-flip scale $\tilde{\w}_{\m}$ (peak maximum) progressively diminishes upon increasing $\tilde{U}$, and $\tilde{\w}_{\m} \ra 0$ as the SC/LM transition is approached.  Inset:  corresponding $\mbox{Im}^0\!\Pi^{+-}(\omega)$ for $\tilde{U} = 13$, showing the high-energy Stoner band.  Note the very different scales in the two figures.}
\end {center}
\end{figure}

We now return to consider $\Sigma_\up(\w)$, equation (5.8).  In the SC phase, $\Pi^{+-}(\w)$ obeys the same Hilbert transform as $\ ^0\!\Pi^{+-}(\omega)$, equation (5.5); using this, together with  
\be
{\cal G}^{\pm}_\sigma(\w)=\mp\int_{-\infty}^\infty\frac{\mbox{d}\w_1}{2\pi \mbox{i}}\ \frac{{\cal G}_\sigma(\w_1)}{\w-\w_1\pm \mbox{i}\eta}
\ee
equation (5.8) reduces to
\be
\fl \ \ \ \Sigma_\up(\w)=U^2\int_{-\infty}^\infty\frac{\mbox{d}\w_1}{\pi}\ \mbox{Im}\Pi^{+-}(\w)\left[\theta(\w_1){\cal G}^-_\down(\w_1+\w)+\theta(-\w_1){\cal G}^+_\down(\w_1+\w)\right].
\ee
Since $\mbox{Im}{\cal G}^{\pm}_\sigma(\w)=\mp\pi D^0_\sigma(\w)\theta(\pm \w)$ (see equation (5.6)), equation (5.12) yields
\alpheqn
\be
\Sigma_\up(\w)=\Sigma^{\subr}_{\up}(\w)-\mbox{isgn}(\w)\Sigma^{\subi}_\up(\w)
\ee
where
\bea
\Sigma^{\subi}_\up(\w)=&\theta(-\w)U^2\int_0^{|\w|}\mbox{d}\w_1\ \mbox{Im}\Pi^{+-}(\w_1)D^0_\down(\w_1+\w)\nonumber \\
&+\theta(\w)U^2\int_{-|\w|}^{0}\mbox{d}\w_1\ \mbox{Im}\Pi^{+-}(\w_1)D^0_\down(\w_1+\w).
\eea
\reseteqn
$\mbox{Im}\Pi^{+-}(\w)$ is given from equation (5.2) by
\be
\mbox{Im}\Pi^{+-}(\w)=\frac{\mbox{Im}\ ^0\!\Pi^{+-}(\omega)}{\left[1-U\mbox{Re}\ ^0\!\Pi^{+-}(\omega)\right]^2+\left[U\mbox{Im}\ ^0\!\Pi^{+-}(\omega)\right]^2}
\ee
and is non-negative (as follows from equation (5.7)); so too is $D^0_\sigma(\w)$ (equation (4.3)).  Hence $\Sigma^{\subi}_\up(\w)\geq0$ as required by analyticity; and $\Sigma^{\subr}_\up(\w)/\Sigma^{\subi}_\up(\w)$ are related by the Hilbert transform, equation (2.7) with $F=\Sigma_\up$.  

To obtain the low-$\w$ behaviour of $\Sigma^{\subi}_\up(\w)$ from equation (5.13b), we require that for $D^0_\down(\w)$ and $\mbox{Im}\Pi^{+-}(\w)$.  For the SC phase, $U\mbox{Re}\ ^0\!\Pi^{+-}(\omega = 0)< 1$ (as discussed further below); hence from equation (5.14), $\mbox{Im}\Pi^{+-}(\w)\propto \mbox{Im}\ ^0\!\Pi^{+-}(\omega)$ as $\w \rightarrow 0$, with $\mbox{Im}\ ^0\!\Pi^{+-}(\omega)$ given by equation (5.7).  In what follows we consider explicitly the case where $|\mu|>0$ self-consistently, for three reasons. (i) This is naturally the case relevant to strong coupling (large $U$) behaviour for any $r\geq 0$; in particular to the SC/LM phase boundary (section 6), to the asymptotic behaviour of the Kondo scale (section 6), and to the consequent universal scaling behaviour of single particle spectra in the SC phase (section 8).  (ii) The case where $|\mu|=0$ (self-consistently) differs only in detail from $|\mu|>0$; the main conclusions reached below hold also for $|\mu|=0$. (iii) We consider the case of $|\mu|=0$ in section 7.1, where we show in particular that as $U \rightarrow 0$ our description of the SC phase is perturbatively exact to (and including) second order in $U$ about the non-interacting limit.

From equation (5.7), using equation (4.5) for $D^0_\sigma(\w)$, the low-$\w$ asymptotic behaviour of $\mbox{Im}\ ^0\!\Pi^{+-}(\omega)$ is
\be
\mbox{Im}\ ^0\!\Pi^{+-}(\omega)\stackrel{\w \rightarrow 0}{\sim}\left[\frac{\Delta_0}{\pi x^2}\right]^2B(r)|\w|^{1+2r}
\ee
where $B(r)=\sqrt{\pi}\Gamma(1+r)/\left[2^{1+2r}\Gamma(\frac{3}{2}+r)\right]$; hence
\be
\mbox{Im}\Pi^{+-}(\w)\propto |\w|^{1+2r}\ \ \ \ \ :\w \rightarrow 0.
\ee
The low-$\w$ behavior of $\Sigma^{\subi}_\up(\w)$ then follows from equation (5.13b), viz
\be
\left(\tilde{\Sigma}^{\subi}_\up(\w)=\right)\ \ \Sigma^{\subi}_\up(\w) \propto |\w|^{2+3r} \ \ \ \ \  :\w \rightarrow 0
\ee
with a prefactor independent of whether $\w \rightarrow 0+$ or $0-$; and from the Hilbert transform equation (2.7),
\alpheqn
\be
\Sigma^{\subr}_\up(\w)\stackrel{\w \rightarrow 0}{\sim}\Sigma^{\subr}_\up(\w =0)-\gamma \w
\ee
with $\gamma=-\left(\partial \Sigma^{\subr}_\up(\w)/\partial \w \right)_{\w=0}$ given by
\be
\gamma = \int_{-\infty}^{\infty}\frac{\mbox{d}\w}{\pi}\ \frac{\Sigma^{\subi}_\up(\w)}{\w^2} \ \ \ \ \  >0.
\ee
\reseteqn
The full interaction self-energy $\tilde{\Sigma}_\up(\w)=\tilde{\Sigma}^{\subr}_\up(\w)-\mbox{isgn}(\w)\tilde{\Sigma}^{\subi}_\up(\w)$ is given by $\tilde{\Sigma}_\up(\w)=-\frac{1}{2}U|\mu|+\Sigma_\up(\w)$ (equation (3.8)), so the low-$\w$ behaviour of $\tilde{\Sigma}^{\subi}_{\up}(\w)$ is again given by equation (5.17); while if equation (5.9) is satisfied, the asymptotic behaviour of $\tilde{\Sigma}^{\subr}_\up(\w)$ is given from equation (5.18) by
\be
\tilde{\Sigma}^{\subr}_\up(\w)\stackrel{\w \rightarrow 0}{\sim}-\gamma\w.
\ee
Both $\tilde{\Sigma}^{\subr}_\up(\w)$ and $\tilde{\Sigma}^{\subi}_\up(\w)$ thus decay to zero as $\w \rightarrow 0$ more rapidly than $\Delta_{\subi/\subr}\sim |\w|^r$ for any $r<1$.  These are the requisite conditions upon $\tilde{\Sigma}_\sigma(\w)$ for a SC state to arise for $U>0$, as discussed in section 3.  Hence, {\it if} the basic self-consistency equation (5.9) admits a solution, this is a sufficient condition for a SC state.

An important characteristic of the SC state follows directly from the above analysis, namely that as $\w \rightarrow 0$
\be
\tilde{\Sigma}_\up(\w)=\tilde{\Sigma}_\down(\w)=\Sigma(\w)\ \ \ \ \ :\w \rightarrow 0.
\ee
That $\tilde{\Sigma}_\up(\w)=\tilde{\Sigma}_\down(\w)$ to leading order follows directly from equations (5.17,19) using the basic symmetry equation (3.5); from this, using equations (3.2,3) for $G(\w)$ together with the definition equation (2.12) of the single self-energy $\Sigma(\w)$, it follows that $\tilde{\Sigma}_\sigma(\w)=\Sigma(\w)$ as $\w \rightarrow 0$.  Equations (5.17,19) thus give the low-$\w$ behaviour of $\Sigma(\w)$, and encompass as a special case the Fermi liquid behaviour characteristic of the normal ($r=0$) Anderson model, viz $\Sigma^{\subi}(\w)\propto \w^2$.  As will be shown in section 5.2, the above behaviour is in marked contrast to that characteristic of the LM phase.

\subsubsection{SC state: stability condition}

Before turning to the self-energy in the LM phase (section 5.2), and to provide continuity to that discussion, we comment on the evolution in the SC phase of the local moment $|\mu|$ determined self-consistently from equation (5.9); and its associated implications for the low-energy spin-flip scale $\w_{\m}$ illustrated in Figure 4.

\tolerance=20
From the Hilbert transform equation (5.5) appropriate to $\Pi^{+-}(\w)$ in the SC phase, it follows that
\be
\mbox{Re}\Pi^{+-}(\w=0)=\int_{-\infty}^{\infty}\frac{\mbox{d}\w}{\pi}\ \frac{\mbox{Im}\Pi^{+-}(\w)}{|\w|}\ \ \ \ \  >0
\ee
which is positive definite since $\mbox{Im}\Pi^{+-}(\w)\geq 0$.  But $\Pi^{+-}(\w)$ is given by equation (5.2) whence, since $\mbox{Im}^0\!\Pi^{+-}(\omega=0)=0$, $\mbox{Re}\Pi^{+-}(\w=0)=\mbox{Re}^0\!\Pi^{+-}(\omega=0)/\left(1-U\mbox{Re}^0\!\Pi^{+-}(\omega=0)\right)$.  For the stability condition equation (5.21) to be satisfied, $0<U\mbox{Re}^0\!\Pi^{+-}(\omega=0)<1$ is thus required.  And an explicit expression for $\mbox{Re}^0\!\Pi^{+-}(\omega=0)$ is readily deduced from equation (5.3), using equation (5.6) together with the identity ${\cal G}_\up(\w)-{\cal G}_\down(\w) = -U|\mu|{\cal G}_\up(\w){\cal G}_\down(\w)$; namely
\alpheqn
\bea
U\mbox{Re}\ ^0\!\Pi^{+-}(\omega=0)&=\frac{1}{|\mu|}\int_{-\infty}^0\mbox{d}\w\left[D^0_\up(\w)-D^0_\down(\w)\right]\\
\\ \nonumber
&=\frac{f(x)}{|\mu|}
\eea
\reseteqn
where $f(x)$ ($x=\frac{1}{2}U|\mu|$) is thus defined, and has been introduced in section 4.1.  For stability, $|\mu|>f(x)$ is thus required; and since $f(x)$ may be shown to be a monotonically increasing function of $x=\frac{1}{2}U|\mu|$, saturating to 1 as $x \rightarrow \infty$ (as is physically obvious), the condition $|\mu|=2x/U>f(x)$ thus amounts to
\alpheqn
\be
|\mu|>|\mu_0|
\ee
for any given $U$ in the SC phase; where $|\mu_0|$ is given by
\be
|\mu_0|=f(\mbox{$\frac{1}{2}$}U|\mu_0|)
\ee
\reseteqn
and from equation (4.8) is simply the mean field (UHF) local moment (denoted from now on by $|\mu_0|$).

Hence, for the SC state stability condition equation (5.21) to be satisfied, the local moment $|\mu|$ determined self-consistently via equation (5.9) must exceed the corresponding UHF moment $|\mu_0|$.  This is correctly found in practice upon solution of equation (5.9) using equation (5.8) for $\Sigma_\up(\w)$ (as shown explicitly in Figure 6 below); and is reflected in turn in the spectral density of transverse spin excitations, $\mbox{Im}\Pi^{+-}(\w)$, which (see Figure 4) is characterized by a strong resonance centred on the low-energy spin-flip scale $\w_{\m} >0$.  But on increasing $U$ in the SC phase towards the critical $U_{\cc}(r)$ above which solution to equation (5.9) is no longer possible, the self-consistently determined $|\mu|$ approaches $|\mu_0|$ continuously from above and $\w_{\m} \rightarrow 0$ (Figures 4, 6).  At $U=U_{\cc}(r)$ precisely, $|\mu|=|\mu_0|$ and $\w_{\m} =0$.  This is the transition point: here $U_{\cc}\mbox{Re}\ ^0\!\Pi^{+-}(\omega=0)=1$, and the resonance in $\mbox{Im}\Pi^{+-}(\w)$ becomes an isolated pole at $\w =0$.  The latter is the natural signature of the LM phase, since for a doubly degenerate LM state with finite weight on the impurity there is no energy cost to flip a spin; it persists through the LM phase, where $|\mu|=|\mu_0|$, and to consideration of which we now turn.

\subsection{LM state}
We begin by stating the result for $\Pi^{+-}(\w)$ in the LM phase.  It consists of a continuum contribution, denoted $^{\mbox{\ssz{S}}}\Pi^{+-}(\w)$, and an $\w=0$ pole with poleweight $Q>0$; specifically,
\alpheqn
\be
\Pi^{+-}(\w)=-\frac{Q}{\w+\mbox{i}\eta}+\ ^{\mbox{\ssz{S}}}\Pi^{+-}(\w)
\ee
with
\be
Q = \left[U^2 \left(\frac{\partial \mbox{Re}\ ^0\!\Pi^{+-}(\omega)}{\partial \w}\right)_{\w=0}\right]^{-1}\ \ \ >0.
\ee
\reseteqn
$\mbox{Im}^{\mbox{\ssz{S}}}\Pi^{+-}(\w)$ is given by equation (5.14), and the real/imaginary parts of $^{\mbox{\ssz{S}}}\Pi^{+-}(\w)$ are again related by the Hilbert transform equation (5.5).

The pole contribution arises from equation (5.2) for $\Pi^{+-}(\w)$ because (i) $U\mbox{Re}\ ^0\!\Pi^{+-}(\omega=0)=1$ in the LM phase ($|\mu|=|\mu_0|$), and (ii) the low-$\w$ behaviour of $\mbox{Re}\ ^0\!\Pi^{+-}(\omega)$ is linear in $\w$,
\be
\mbox{Re}\ ^0\!\Pi^{+-}(\omega)\stackrel{\w \rightarrow 0}{\sim}\frac{1}{U}+\w\left(\frac{\partial \mbox{Re}\ ^0\!\Pi^{+-}(\omega)}{\partial \w}\right)_{\w=0}
\ee
where $\pi \left(\partial \mbox{Re}\ ^0\!\Pi^{+-}(\omega)/\partial \w\right)_{\w =0} = \int_{-\infty}^{\infty}\mbox{d}\w\ \mbox{Im}\ ^0\!\Pi^{+-}(\omega)\mbox{sgn}(\w)/\w^2$  is readily shown to be positive.  The low-$\w$ behaviour of $\mbox{Im}\ ^0\!\Pi^{+-}(\omega)$ is again given precisely by equation (5.15), and for any $r>0$ decays to zero as $\w \rightarrow 0$ more rapidly than $\w$.  Hence, from equation (5.2), the $\w \rightarrow 0$ behaviour of $\Pi^{+-}(\w)$ is $-Q/\w$ with $Q$ given by equation (5.24b); the $\delta$-function part of which pole is obtained by an analytical continuation $\w \rightarrow \w+\mbox{i}\eta$, which is unique since $Q>0$ and $\mbox{Im}\Pi^{+-}(\w)\geq 0$ necessarily.  Equation (5.24) thus results.

Using equation (5.24) in equation (5.8) for $\Sigma_\up(\w)$, together with equation (5.11), gives the basic form for $\Sigma_\up(\w)$ in the LM phase:
\be
\Sigma_\up(\w)=QU^2{\cal G}^-_\down(\w)+^{\mbox{\ssz{S}}}\Sigma_\up(\w).
\ee
\tolerance=20
The first term, arising from the pole contribution to $\Pi^{+-}(\w)$, controls the low-$\w$ asymptotics of $\Sigma_\up(\w)$ as shown below.  The second, $^{\mbox{\ssz{S}}}\Sigma_\up(\w)$, is given by equation (5.8) with $\Pi^{+-}\rightarrow\mbox{}^{\mbox{\ssz{S}}}\Pi^{+-}$, and by precisely the same argument used in section 5.1 is given by equation (5.12) with $\mbox{Im}\Pi^{+-}\rightarrow \mbox{Im}^{\mbox{\ssz{S}}}\Pi^{+-}$; hence, as in equation (5.13), $^{\mbox{\ssz{S}}}\Sigma_\up(\w)= \mbox{}^{\mbox{\ssz{S}}}\Sigma^{\subr}_\up(\w)-\mbox{isgn}(\w)^{\mbox{\ssz{S}}}\Sigma^{\subi}_\up(\w)$ with $^{\mbox{\ssz{S}}}\Sigma^{\subi}_{\up}(\w)\geq 0$ given by equation (5.13b).  And since ${\cal G}^-_\down(\w)=\mbox{Re}{\cal G}^-_\down(\w)-\mbox{isgn}(\w)\pi D^0_\down(\w)\theta(-\w)$ it follows that $\Sigma_\up(\w)$ in its entirety is given by equation (5.13a); with $\Sigma^{\subi}_\up(\w) \geq 0$ as required by analyticity, and $\Sigma^{\subr}_\up/\Sigma^{\subi}_\up$ related by the Hilbert transform equation (2.7).  $\Sigma^{\subi}_\up(\w)$ is given explicitly by
\be
\left(\tilde{\Sigma}^{\subi}_\up(\w)=\right)\ \ \Sigma^{\subi}_\up(\w)=\pi QU^2D^0_\down(\w)\theta(-\w)+^{\mbox{\ssz{S}}}\Sigma^{\subi}_\up(\w)
\ee
the low-$\w$ asymptotics of which we now consider.

The $\w \rightarrow 0$ behaviour of $^{\mbox{\ssz{S}}}\Sigma^{\subi}_\up(\w)$ is obtained from equation (5.13b) in parallel to the corresponding analysis of section 5.1.  $\mbox{Im}^{\mbox{\ssz{S}}}\Pi^{+-}(\w)$ is given by equation (5.14), with $\mbox{Im}^0\!\Pi^{+-}(\omega)$ by equation (5.15) and $\mbox{Re}^0\!\Pi^{+-}(\omega)$ by equation (5.25); and since $U\mbox{Re}^0\!\Pi^{+-}(\omega=0)=1$, equation (5.14) thus gives $\mbox{Im}^{\mbox{\ssz{S}}}\Pi^{+-}(\w)\propto |\w|^{2r-1}$ as $\w \rightarrow 0$.  Hence, using $D^0_\down(\w)\sim |\w|^r$ as $\w \rightarrow 0$, equation (5.13b) yields
\be
^{\mbox{\ssz{S}}}\Sigma^{\subi}_{\up}(\w)\propto |\w|^{3r}\ \ \ \w \rightarrow 0.
\ee
But since $D^0_\down(\w)\propto |\w|^r$ as $\w \rightarrow 0$ (equation (4.5)), it is the first term in equation (5.27) that controls the low-$\w$ behaviour of $\Sigma^{\subi}_\up(\w)$ for any $r>0$ (and it is only for $r>0$ that a LM state arises, as will be shown in section 6).  It can moreover be shown that the poleweight $Q$ (equation (5.24b)) is given simply by $Q=|\mu_0|^2$; hence, using equation (4.5) the $\w \rightarrow 0$ behaviour of $\Sigma^{\subi}_\up(\w)=\tilde{\Sigma}^{\subi}_\up(\w)$ in the LM phase is:
\be
\Sigma^{\subi}_\up(\w)\stackrel{\w \rightarrow 0}{\sim}4\Delta_0|\w|^r\theta(-\w).
\ee
$\Sigma^{\subi}_\down(\w)=\Sigma^{\subi}_\up(-\w)$ follows directly by symmetry and we note that in contrast to the SC phase (equation (5.20)), $\Sigma_\up(\w)$ and $\Sigma_\down(\w)$ do {\it not} therefore coincide to leading order as $\w \rightarrow 0$, the physical significance of which will be discussed in section 8.

The corresponding real part $\Sigma^{\subr}_\up(\w)$, and hence $\tilde{\Sigma}^{\subr}_\up(\w)=-\frac{1}{2}U|\mu_0|+\Sigma^{\subr}_\up(\w)$, follows by Hilbert transformation.  Here we simply note that $\Sigma^{\subr}_\up(\w=0)<\frac{1}{2}U|\mu_0|$ is found throughout the LM phase whence, in contrast to the SC phase (equation (5.9)), $\tilde{\Sigma}^{\subr}_\up(\w=0)<0$.  This, together with equation (5.29), enables the low-$\w$ behaviour of the full single-particle spectrum $D(\w)=-\pi^{-1}\mbox{sgn}(\w)\mbox{Im}G(\w)$ to be obtained; for from equations (3.2,3), $D(\w)=\frac{1}{2}\left[D_\up(\w)+D_\down(\w)\right]$ with $D_\sigma(\w)$ given as $\w \rightarrow 0$ by
\alpheqn
\be
D_\sigma(\w)\stackrel{\w \rightarrow 0}{\sim}\frac{\left[\Delta_{\subi}(\w)+\Sigma^{\subi}_\sigma(\w)\right]}{\pi\left[\tilde{\Sigma}^{\subr}_\sigma(\w=0)\right]^2}.
\ee
And since $\Sigma^{\subr}_\down(\w=0)=-\Sigma^{\subr}_\up(\w=0)$, it follows directly using equation (5.29) that
\be
D(\w)\stackrel{\w \rightarrow 0}{\sim}\frac{3\Delta_0}{\pi\left[\tilde{\Sigma}^{\subr}_\up(\w=0)\right]^2}|\w|^r.
\ee
\reseteqn

The characteristic low-$\w$ spectral signature of the LM phase found in NRG calculations [28], $D(\w)\propto |\w|^r$, is thus recovered; and we add that equation (5.30b) holds for the LM state regardless of whether $r \lessgtr 1$.  For the particular case of $r>1$ however, equation (5.30b) is readily shown to be asymptotically exact as $U \ra 0$.  Here, $\tilde{\Sigma}^{\subr}_\up(0)$ is dominated by the Fock contribution of $-\frac{1}{2}U|\mu_0|$; and as discussed in section 4 (equation (4.13) and Appendix), $|\mu_0| \ra q$ as $U \ra 0$, with $q$ (equation (2.10b)) the weight of the $\w=0$ pole in the non-interacting single-particle spectrum $d_0(\w)$.  Hence as $U\ra 0$:
\be
D(\w)=\stackrel{\w \ra 0}{\sim}\frac{12}{\pi}\frac{\Delta_0}{(Uq)^2}|\w|^r\ \ \ \ \ :U \ra 0,\ r>1.
\ee
This is precisely the result obtained by us hitherto (equation (5.12) of [29]) using straight second-order perturbation theory in $U$ about the non-interacting limit, which is itself applicable for $r>1$ (but not for $\frac{1}{2}<r<1$ [29]).  Note that this result is not captured correctly at pure mean-field level alone which, from equation (4.5) with $|\mu_0|=q$, differs by a factor of 3 (the presence of which reflects the fact that equation (5.29) for $\Sigma^{\subi}_\up(\w)$ as $\w \ra 0$ is independent of $U$).  That equation (5.31) is correctly recovered as a limiting case of the present theory is thus a non-trivial consequence of the local moment approach.

Finally, the low-$\w$ behaviour in the LM phase of the conventional single self-energy $\Sigma(\w)$ --- defined by equation (2.12) --- may also be deduced from the above asymptotics.  We consider explicitly $r<1$ (the case $r>1$ will be discussed in section 7.1).  Consider the Hilbert transform equation (2.7) for $F(\w)=F_{\subr}(\w)-\mbox{isgn}(\w)F_{\subi}(\w)$.  If the $\w \ra 0$ behaviour of $F_{\subi}(\w)$ is
\alpheqn
\be
F_{\subi}(\w)\stackrel{\w \ra 0}{\sim}\alpha |\w|^\lambda
\ee
with $-1<\lambda<1$, the low-$\w$ behaviour of $F_{\subr}(\w)$ is readily shown to be
\be
F_{\subr}(\w)\stackrel{\w \ra 0}{\sim}-\mbox{sgn}(\w)\mbox{tan $\left(\frac{\pi}{2}\lambda \right)$}F_{\subi}(\w).
\ee
\reseteqn
With $F_{\subi}(\w)\equiv \pi D(\w)$ given as $\w \ra 0$ by equation (5.30b), $\mbox{Re}G(\w)\equiv F_{\subr}(\w)$ thus follows directly; and from equation (2.12),  $\Sigma(\w)=\w+\mbox{i}\eta \mbox{sgn}(\w)-\Delta(\w)-G^{-1}(\w)$ is in consequence given asymptotically by
\alpheqn
\be
\Sigma^{\subr}(\w)\stackrel{\w \ra 0}{\sim}\mbox{sgn}(\w)\mbox{tan $\left(\frac{\pi}{2}r\right)$}\Sigma^{\subi}(\w)
\ee
with
\be
\Sigma^{\subi}(\w)\stackrel{\w \ra 0}{\sim}\frac{\left[\tilde{\Sigma}^{\subr}_\up(\w=0)\right]^2}{3\Delta_0}\mbox{cos$^2\left(\frac{\pi}{2}r\right)$}|\w|^{-r} \ \ : \ \ r<1
\ee
\reseteqn
This divergent behaviour of $\Sigma(\w)$ as $\w \ra 0$ in the LM phase --- which as just seen is a direct consequence of $D(\w) \propto |\w|^r$ as $\w \ra 0$ --- is in marked contrast both to $\Sigma_\sigma(\w)$ (equation (5.29)); and to the behaviour of $\Sigma(\w)$ in the SC phase (equations (5.17,19,20)) where, as befits a generalized Fermi-liquid state, $\Sigma^{\subi}(\w)$ vanishes at the Fermi level, $\w =0$.
\seceq
\section{Statics}
We now consider the ramifications of the local moment approach (LMA) developed in the preceeding sections, beginning with `statics'; dynamics, in the form of single-particle excitation spectra, will be investigated in sections 7 and 8.  Specifically, we consider here: (i) the phase boundaries between SC and LM states, including their predicted scaling behaviour (section 6.3) and quantitative comparison with NRG results [21,28] (section 6.4).  As mentioned in section 4, the problem is characterized generally by two dimensionless material parameters: the reduced interaction strength, $\tilde{U}=U/\delr$, and bandwidth $\tilde{D}=D/\delr$; or, equivalently, by $\tilde{U}$ and $U/D$.  We seek the critical $\tilde{U}_{\cc}(r)$ versus $r$ phase boundaries, as a function of $U/D$.  (ii) The evolution, and critical behaviour, of the central low-energy spin-flip (or Kondo) scale, $\w_{\m}$, that is symptomatic of the SC state and was discussed briefly in section 5.1 (see Figure 4); this is considered in section 6.1. (iii) The relationship (section 6.3) between the soft-gap Anderson model and the corresponding Kondo model in the strong coupling (large $U$) regime [21,28].  Particular attention will be given throughout to the behaviour of physical properties at low $r$, which is of evident importance in connecting to the `normal' ($r=0$) Anderson model; and for which the salient results, of which the most important are asymptotically exact, can be extracted analytically.

To motivate our subsequent discussion, we begin with a phase diagram obtained from the LMA as detailed in section 5.1, via the limits of solution to equation (5.9) appropriate to the $U>0$ SC state.  Figure 5 shows the resultant critical $\Delta_0U^r/U (=\tilde{U}^{r-1})$ versus $r$, obtained strictly for the wide band limit $D=\infty$ (but, as shown in section 6.3, coincident in practice with that for $U/D \lesssim 0.1$); the inset shows the low $r<0.1$ behaviour on an expanded scale, and the main figure includes for comparison the static mean-field (UHF) result obtained in section 4.1 (Figure 1).

\begin{figure}
\begin{center}
\epsfig{file =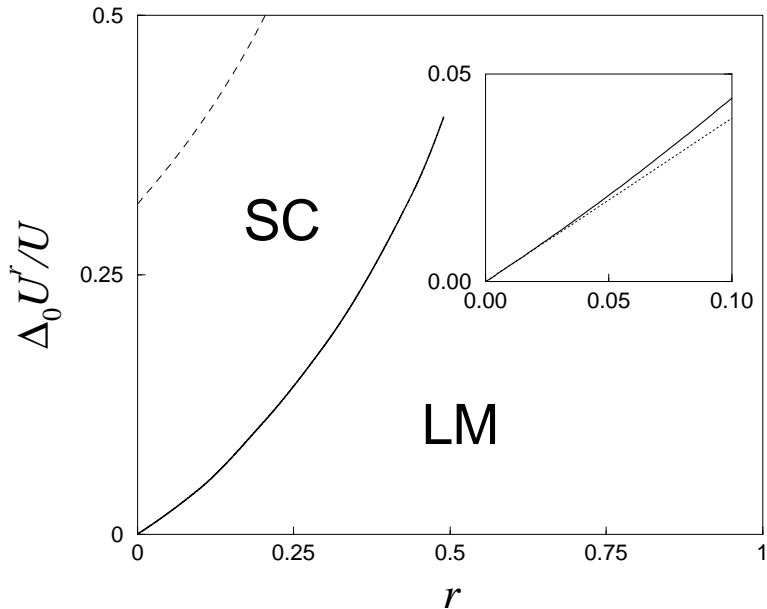,width=10cm} 
\caption{LMA phase diagram, $(\Delta_0U^r/U)_{\cc}=\tilde{U}_c^{r-1}$ versus $r$ (wide band limit); for $r>\frac{1}{2}$, solely LM states occur for all $U>0$.   The mean-field phase boundary is shown for comparison (dashed line).  Inset: $r<0.1$ behaviour on expanded scale including the $r \ra 0$ Kondo asymptote (dotted) of $\pi r/8$ (see section 6.1, equation (6.10)).}
\end {center}
\end{figure}

As seen from Figure 5, the small-$r$ behaviour of the phase boundary differs radically from that of UHF, which yields a spurious SC/LM transition even for $r=0$.  The LMA phase boundary, by contrast, vanishes linearly in $r$ as $r \ra 0$, as found in NRG studies of both the Anderson [21,28] and Kondo [22,27] models; the precise form of $\tilde{U}_{\cc}(r)$ as $r \ra 0$ will be established in section 6.1.  For the normal $r=0$ Anderson model in particular, the LMA correctly recovers $(\Delta_0/U)_{\cc}=0$: the LM phase is here confined entirely to the atomic limit $\Delta_0=0$, and for all $\Delta_0>0$ the system is a normal Fermi liquid with a strong coupling Kondo scale whose asymptotics within the LMA will be obtained analytically in section 6.1 from the limit $r \ra 0$, and shown to coincide with that arising from scaling and the Bethe ansatz.  Upon increasing $r$ from zero the critical $\Delta_0U^r/U$ departs quite rapidly from linearity, turns upward and then terminates at $r=\frac{1}{2}$.  The latter behaviour is different from that arising at mean-field level (where the boundary curve diverges as $r \ra \frac{1}{2}-$, Figure 1), but has been reported in an NRG study of the Kondo model [27]; we return to this issue when making detailed comparison with NRG results in section 6.4.

One small but important point should be noted regarding the LMA phase boundaries: the same $\tilde{U}_{\cc}(r)$ is correctly obtained whether one approaches the boundary from the SC phase ($\tilde{U}<\tilde{U}_{\cc}(r)$) or the LM phase ($\tilde{U}>\tilde{U_{\cc}}(r)$).  Recall the statement above that $\tilde{U}_{\cc}(r)$ is obtained from the limit of solutions to equation (5.9) --- $\tilde{\Sigma}^{\subr}_\up(\w=0)=0$ --- appropriate to the $U>0$ SC state; such that for $\tilde{U}>\tilde{U}_{\cc}(r)$, $\tilde{\Sigma}^{\subr}_\up(0)=0$ cannot be satisfied self-consistently.  As mentioned in section 5.2, we find by contrast that $\tilde{\Sigma}^{\subr}_\up(0)<0$ throughout the LM phase.  For the same $\tilde{U}_{\cc}(r)$ to arise coming from the LM phase, $\tilde{U}>\tilde{U}_{\cc}(r)$, thus requires $\tilde{\Sigma}^{\subr}_\up(0) \ra 0-$ as $\tilde{U} \ra \tilde{U}_{\cc}(r)+$ (on the assumption of continuity).  This is indeed precisely as found in practice.

The final qualitative feature of the phase diagram shown in Figure 5 is that for $r>\frac{1}{2}$ solely LM states occur for all $U>0$, as is indeed found from NRG studies [21,28] and seen already at mean-field level (Figure 1).  Here we find, consistently, that $\tilde{\Sigma}^{\subr}_\up(\w =0)<0$ for all finite $\tilde{U}$, and vanishes only as $\tilde{U} \ra 0$ (where the interaction self-energy is of course zero by construction).

\subsection{Strong coupling asymptotics}
We now consider the asymptotic behaviour of the SC phase in the spin-fluctuation regime of strong-coupling (meaning large $U$).  From the phase diagram Figure 5, this formally entails consideration of the important limiting behaviour $r \ra 0$ (where $\tilde{U}_{\cc}(r) \ra \infty$), since charge fluctuations can strictly be neglected only as $\tilde{U} \ra \infty$; although in practice the results obtained below naturally hold over a finite-$r$ range.  Our aim is to obtain the critical $\tilde{U}_{\cc}(r)$ and, relatedly, the behaviour of the low energy spin-flip or Kondo scale $\w_{\m}(r)$ as $\tilde{U}\ra \tilde{U}_{\cc}(r)-$ from the SC phase.  No restriction is imposed on the bandwidth, $D$, and the impurity level $\epsilon_i=-U/2$ may lie within or outside the band; to encompass which we define $\lambda$, used in the following, by
\be
\lambda=\mbox{min}\left[D,\mbox{$\frac{U}{2}$}\right].
\ee

The key to extracting the strong coupling asymptotics naturally lies in $\Sigma_\up(\w)$, given for the SC phase by equation (5.12); and in particular in $\Sigma^{\subr}_\up(\w=0)$, whose large-$U$ form is readily deduced from two properties of the transverse spin polarization propagator $\mbox{Im}\Pi^{+-}(\w)$ (itself illustrated in Figure 4).  First, that in strong coupling the spectral weight of $\mbox{Im}\Pi^{+-}(\w)$ is confined entirely to frequencies $\w>0$ (as already evident in Figure 4), and in consequence the first term on the right hand side of equation (5.12) for $\Sigma^{\subr}_\up(\w=0)$ is dominant in determining its asymptotics.  Specifically one finds
\be
\int_0^{\infty}\frac{\mbox{d}\w}{\pi}\ \mbox{Im}\Pi^{+-}(\w)\ra 1
\ee
in strong coupling, which behaviour reflects physically the saturation of the local moment, $|\mu| \ra 1$: it is straightforward to show, and physically rather obvious, that the moment saturates for $\Delta_{\subi}(\w = \lambda)/|\epsilon_i| \ll 1$; i.e.
\be
\frac{U}{2}\gg \Delta_0\lambda^r.
\ee
Second, as illustrated in Figure 4, the strong resonance in $\mbox{Im}\Pi^{+-}(\w)$ occurs on the low-energy spin-flip scale, $\w_{\m}$, that diminishes rapidly with increasing $\tilde{U}$ and vanishes as $\tilde{U}\ra \tilde{U}_{\cc}(r)-$ from the SC phase.  $\mbox{Im}\Pi^{+-}(\w)$ is thus in practice non-zero only on the scale of $\w_{\m}$; and on scales of this order $\mbox{Re}{\cal G}^-_\down(\w)$ is a slowly varying function of frequency.  Hence, the strong coupling behaviour of $\Sigma^{\subr}_\up(\w=0)$ is given asymptotically from equation (5.12) by
\bea
\Sigma^{\subr}_\down(\w=0)\sim& U^2 \mbox{Re}{\cal G}^-_\down(\w_{\m})\int_{0}^\infty \frac{\mbox{d}\w_1}{\pi}\ \mbox{Im}\Pi^{+-}(\w_1)\\ \nonumber
\\ \nonumber
&\ \ \ =U^2\mbox{Re}{\cal G}^-_\down(\w_{\m}).
\eea

The $U$-dependence of $\w_{\m}(r)$ --- and in consequence the critical $U_{\cc}(r)$ --- can now be determined from equation (5.9), which (as discussed in section 5.1) ensures that the generalized spectral pinning condition characteristic of the SC state, equation (2.16), is satisfied; for in strong coupling, where $|\mu|\ra 1$, equation (5.9) reduces using equation (6.4) to
\be
U^2\mbox{Re}{\cal G}^-_\down(\w_{\m})=\frac{1}{2}U.
\ee
This can be solved for $\w_{\m}(r)$ once $\mbox{Re}{\cal G}^-_\down(\w)$ is known, and to which we now turn.

$\mbox{Re}{\cal G}^-_\down(\w)$ is given from equation (5.6) by the one-sided Hilbert transform
\be
\mbox{Re}{\cal G}^-_\down(\w)=\int_{-\infty}^0 \mbox{d}\w_1 D^0_\down(\w_1)\mbox{P}\left(\frac{1}{\w -\w_1}\right)
\ee
whose $\w \ra 0$ behaviour is dominated by that of $D^0_\down(\w_1)$ as $\w_1 \ra 0$, given generally by $D^0_\down(\w_1)\sim (\Delta_0/\pi x^2)|\w_1|^r$ (equation (4.5)) with $x = \frac{1}{2}U|\mu|$; and from the latter alone it can be shown that the asymptotic behaviour of $\mbox{Re}\left[{\cal G}^-_\down(\w)-{\cal G}^-_\down(0)\right]$ is given exactly for $r<1$ by
\be
\mbox{Re}\left[{\cal G}^-_\down(\w)-{\cal G}^-_\down(0)\right]\stackrel{\w \ra 0+}{\sim}-\frac{\Delta_0}{x^2}\frac{1}{\mbox{sin $\left(\pi r\right)$}}|\w|^r +\mbox{O}(|\w|)
\ee
 --- although this by itself does not give $\mbox{Re}{\cal G}^-_\down(\w=0)$.  An approximation to $\mbox{Re}{\cal G}^-_\down(\w)$ that correctly captures this asymptotic behaviour is
\alpheqn
\be
\mbox{Re}{\cal G}^-_\down(\w)\stackrel{\w \ra 0}{\sim}\frac{\Delta_0}{\pi x^2}\int_{-\lambda}^0\mbox{d}\w_1\ |\w_1|^r\mbox{P}\left(\frac{1}{\w -\w_1}\right)
\ee
where we have introduced a high-energy cut-off of order $\lambda=\mbox{min}[D,\frac{U}{2}]$.  We emphasize that the specific cut-off used is wholly inessential to the following arguments: the important point is that the prefactor to the integral is precisely $\Delta_0/\pi x^2$. Evaluation  of equation (6.8a) for $\w>0$ gives
\be
\mbox{Re}{\cal G}^-_\down(\w)\stackrel{\w \ra 0}{\sim}\frac{\Delta_0}{\pi x^2}\left(\frac{\lambda^r}{r}-\frac{\pi}{\mbox{sin($\pi r$)}}|\w|^r\right)+\mbox{O}(|\w|/\lambda)
\ee
\reseteqn
use of which in equation (6.5), with $x\equiv\frac{1}{2}U$ in strong coupling ($|\mu| \ra 1$), yields
\be
U=\frac{8\Delta_0}{\pi}\left(\frac{\lambda^r}{r}-\frac{\pi}{\mbox{sin($\pi r$)}}\ \w_{\m}^r\right).
\ee

The critical $U_{\cc}(r)$, where $\w_{\m}$ vanishes, follows immediately from
\alpheqn
\be
\left(\frac{\Delta_0 \lambda^r}{U}\right)_{\cc}=\frac{\pi r}{8}
\ee
(and of course holds asymptotically as $r \ra 0$ where $U_{\cc}(r) \ra \infty$); equivalently, in the reduced units $\tilde{U}=U/\delr$ and $\tilde{D}=D/\delr$,
\be
\mbox{$\left(\frac{\left[\mbox{min}\left(\tilde{D},\frac{\tilde{U}}{2}\right)\right]^r}{\tilde{U}}\right)_{\cc}$}=\frac{\pi r}{8}
\ee
\reseteqn
yielding in either case
\be
\tilde{U}_{\cc}(r)\sim \frac{8}{\pi r}
\ee
as $r \ra 0$.  The behaviour equation (6.10) is evident in Figure 5 (inset) appropriate to the wide band limit, where the critical $\Delta_0U^r/U=\tilde{U}^{r-1}$ determined by full numerical solution of equation (5.9) is compared to the small-$r$ result $\pi r/8$.  The latter is indeed seen to be asymptotically approached as $r \ra 0$, which behaviour is reached in practice for $r \lesssim 0.02$ (a feature we compare to NRG calculations in section 6.4); we also add in passing that the asymptotic region of linearity in $r$ widens upon introduction of a  finite bandwidth $D$, as will be shown in section 6.3.

For $U<U_{\cc}(r)$ in the SC phase, the spin-flip or Kondo scale $\w_{\m}$ likewise follows directly from equation (6.9), being given as $r \ra 0$ by
\alpheqn
\bea
\w_{\m}(r)\stackrel{r \ra 0}{\sim}&\lambda\left[1-\frac{U}{U_{\cc}(r)} \right]^{\frac{1}{r}}\\
&=\lambda \left[1-r\frac{\pi U}{8 \Delta_0 \lambda^r}\right]^{\frac{1}{r}}
\eea
\reseteqn
and vanishing as $U \ra U_{\cc}(r)-$ with the characteristic exponent $1/r$ for $r>0$; while the limit of $r=0$ yields
\be
\w_{\m}(r=0)=\mbox{min}\left[D,\mbox{$\frac{U}{2}$}\right]\mbox{exp}\left(-\frac{\pi U}{8 \Delta_0}\right)
\ee
which is the Kondo scale characteristic of the normal Anderson model and is exponentially small in strong coupling.  The prefactor, $\lambda=\mbox{min}\left[D,\frac{U}{2}\right]$, merely reflects the high-energy cut-off used in equation (6.8a); but the exponent of $-\pi U/8 \Delta_0$ is exact, agreeing precisely with the result obtained from the Bethe ansatz [34] for the wide-band limit of the $r=0$ Anderson model and, more generally, with poor man's scaling [33] (see e.g. [2]).  Note moreover that recovery of this exponent hinges on the asymptotic validity of equation (6.10) for the phase boundary as $r \ra 0$, which we likewise believe to be exact.

Finally, while the results above are strictly valid as $r \ra 0$, the exponent of $1/r$ for the vanishing of the Kondo scale $\w_{\m}(r)$ as $U \ra U_{\cc}(r)-$, is found to hold generally within the present LMA.  Figure 6 shows the numerically determined $\tilde{\w}_{\m} = \w_{\m}/\delr$ verus $\tilde{U}$ for $r=0.2$ (in the wide-band limit); and as $\tilde{U} \ra \tilde{U}_{\cc}- \simeq 15.8$, careful numerical analysis shows $\w_{\m} \sim \left[1-\tilde{U}/\tilde{U}_{\cc}\right]^5$.  For $\tilde{U}>\tilde{U}_{\cc}$ in the doubly degenerate LM phase, $\w_{\m} = 0$ (as indicated in Figure 6); but is nonetheless `present', giving rise to the $\w =0$ pole in $\mbox{Im}\Pi^{+-}(\w)$ throughout the LM phase (section 5.2) and in consequence to the leading term $QU^2{\cal G}^-_\down(\w)$ contributing to the LM self-energy $\Sigma_\up(\w)$, equation (5.26) (whose existence is in turn responsible for e.g. the asymptotically exact result equation (5.31)).  Figure 6 also shows the evolution of the self-consistently determined local moment $|\mu|$ as $\tilde{U}$ is increased towards and through $\tilde{U}_{\cc}(r)$, together with its mean-field counterpart $|\mu_0|$, and illustrates the discussion of section 5.1.1; specifically that $|\mu|>|\mu_0|$ in the SC phase, with $|\mu|\ra |\mu_0|+$ as $\tilde{U} \ra \tilde{U}_{\cc}-$.
\begin{figure}
\begin{center}
\epsfig{file =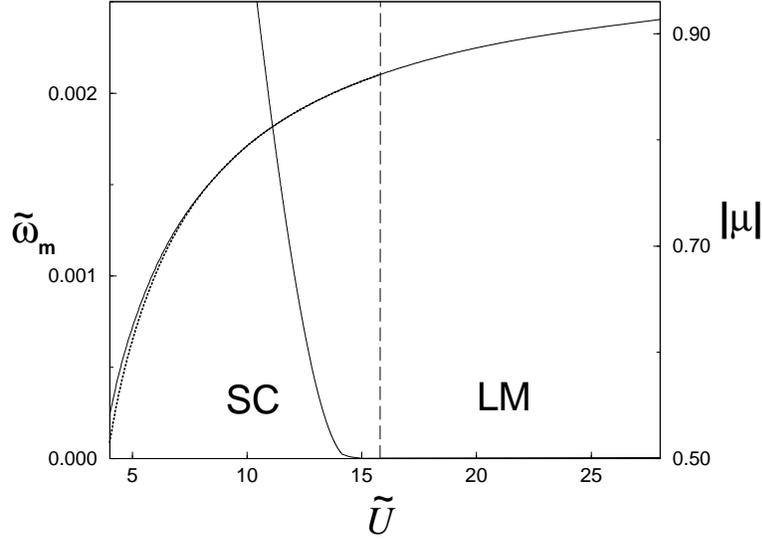,width=10cm} 
\caption{For $r=0.2$, the spin-flip/Kondo scale $\tilde{\w}_{\m}=\w_{\m}/\delr$ versus $\tilde{U}=U/\delr$ (left-hand scale) in the vicinity of the SC/LM transition; and the local moment $|\mu|$ versus $\tilde{U}$ (right scale, solid line), compared to the mean field $|\mu_0|$ (dotted line).  As $\tilde{U}\ra \tilde{U}_{\cc}-$, $\tilde{\w}_{\m} \sim [1-\tilde{U}/\tilde{U}_{\cc}]^5$; and $\tilde{\w}_{\m}=0$ in the LM phase, $\tilde{U}>\tilde{U}_{\cc}$.}
\end {center}
\end{figure}

\subsection{Connection to the Kondo model}
We comment briefly on the connections to the soft-gap Kondo model, which itself has been studied extensively by NRG [26,27], poor man's scaling [22] and large-$N$ mean-field methods [22]; and to which the results of the previous section are clearly related: equation (6.3) is just the condition for applicability of the Schrieffer-Wolff transformation [32] mapping the Anderson to the Kondo model; and from equation (6.10a) is manifestly satisfied for $r\ll 1$ as the SC/LM phase boundary is approached.

The Kondo Hamiltonian $\hat{H}_{\mbox{\ssz{K}}}=\hat{H}_{\mbox{\ssz{host}}}+\hat{H}_{s-d}$ consists of the host band contribution (equation (2.1)), with spectral density
\be
\rho_{\mbox{\ssz{host}}}(\w)=C|\w|^r\theta(D-|\w|);
\ee
and the $s-d$ interaction $\hat{H}_{s-d}=\frac{1}{2}J\sum_{\bi{k,k'}}\left[c^+_{\bi{k}\up}c_{\bi{k'}\down}\hat{S}^-_i+c^+_{\bi{k}\down}c_{\bi{k'}\up}\hat{S}^+_i+\left(c^+_{\bi{k}\up}c_{\bi{k'}\up}-c^+_{\bi{k}\down}c_{\bi{k'}\down} \right)\hat{S}_{iz}\right]$ with an exchange coupling constant, $J$, given from the Schrieffer-Wolff transformation in the particle-hole symmetric case by $J=8V^2/U$ (where a constant $V_{i\bi{k}}=V$ is taken).  From equations (2.5,6) and (6.14), $V^2$ is related to the hybridization parameter of the Anderson model by $\Delta_0=\pi V^2C$, whence
\be
J = \frac{8}{\pi C}\frac{\Delta_0}{U}.
\ee

The critical value of $\Delta_0/U$ at the SC/LM phase boundary is however given within the LMA by equation (6.10a) as $r \ra 0$; hence as $r \ra 0$ the critical $J_{\cc}(r)$ in the Kondo model is
\be
J_{\cc}(r)\stackrel{r \ra 0}{\sim}\frac{r}{C \lambda^r}
\ee
with $\lambda = \mbox{min}\left[D,\frac{U}{2}\right]$.  This is precisely the result for the Kondo model obtained by Withoff and Fradkin [22] from poor man's scaling with $D$ finite, and hence $\lambda=D$.  Likewise equation (6.12b), which with $\lambda=D$ may be cast as $\w_{\m}(r)=D\left(1-J_{\cc}(r)/J\right)^{\frac{1}{r}}$, recovers precisely the Kondo scale obtained by Whitoff and Fradkin via a large-$N$ mean field treatment [22] (and denoted therein by $T_0$).  The LMA for the soft-gap Anderson model thus recovers precisely the correct asymptotics of the corresponding Kondo model in the limit $r \ra 0$.

\subsection{Phase boundaries: scaling}
We return now to the Anderson model for arbitrary $r$, to discuss the scaling characteristics of the phase boundaries predicted by the LMA, and illustrated in Figures 7 and 8.  Figure 7 shows the critical $\Delta_0U^r/U (=\tilde{U}^{r-1})$ versus $r$ curves for five values of $U/D$: $10^{-3},\ 10^{-2},\ 1,\ 10$ and 100; from which is seen that for $U/D \ll 1$ (and in practice for $U/D \lesssim$0.1--1) the phase boundaries collapse to a common curve, namely the wide-band limit shown in Figure 5.  By contrast, and for the same $U/D$ values, Figure 8 shows the resultant critical $\Delta_0D^r/U (=\tilde{D}^r/\tilde{U})$ versus $r$ curves; from which it is evident that for $U/D \gg 1$ (and in practice $U/D\gtrsim$ 10) common scaling of the phase boundaries again arises.  Thus, although the model contains two independent parameters --- viz $\tilde{U}=U/\delr$ and $U/D$ --- it is clear that the phase boundaries exhibit one-parameter scaling for the two distinct regimes $U/D \ll1$ and $\gg1$, according to whether the impurity level $|\epsilon_i|=U/2$ lies respectively well within or outside the band.

\begin{figure}
\begin{center}
\epsfig{file =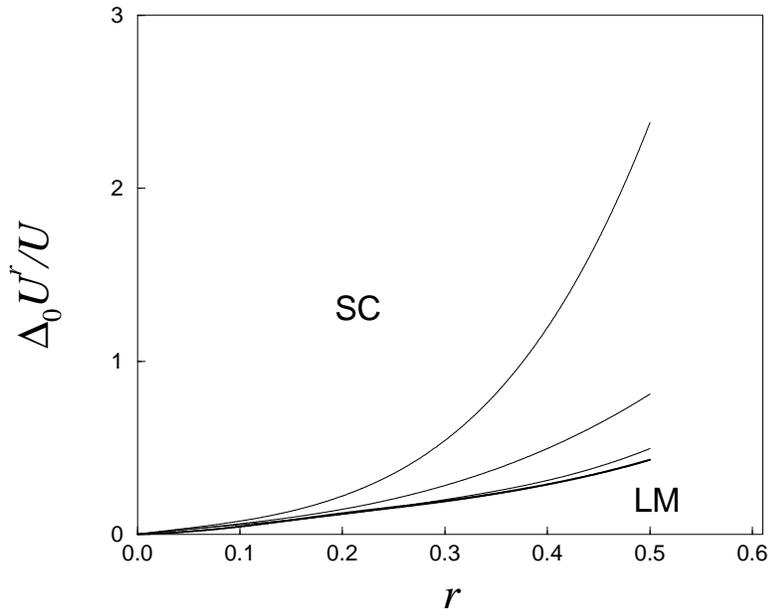,width=10cm} 
\caption{LMA phase boundaries $(\Delta_0U^r/U)_{\cc}$ versus $r$ for (top to bottom) $U/D$ = 100, 10, 1, $10^{-2}$ and $10^{-3}$; for $U/D \ll 1$ common scaling occurs (bold line), as discussed in text.}
\end {center}
\end{figure}

The behaviour found generally above is in fact suggested by the asymptotic low-$r$ result of section 6.1, viz $(\Delta_0\lambda^r/U)_{\cc}=\pi r/8$ with $\lambda = \mbox{min}\left[D,\frac{U}{2}\right]$.  Moreover, the scaling illustrated in Figure 7 for $U\ll D$ has been observed in NRG calculations by Gonzalez-Buxton and Ingersent [21], who find excellent scaling for $U/D = 0.2$ and $0.02$ (see Figure 20 of [21], where for the symmetric model the quantity $\rho_0J_{\cc}$ there plotted is precisely $(8/\pi)(\Delta_0 \lambda^r/U)_{\cc}$ with $\lambda = U/2$).  These authors have also given an argument, based upon poor man's scaling for the soft-gap Anderson model [23], as to why $(\Delta_0U^r/U)_{\cc}\equiv \tilde{U}_{\cc}^{r-1}$ should be universal; although we note that a simpler argument explains this: as the bandwidth $D \ra \infty$ --- which is possible for $r<1$ (section 2.1) --- this scale in effect drops out of the problem, which thus depends solely on the dimensionless ratio $\tilde{U}=U/\delr$.

For $U/D \gg 1$ by contrast, both $\tilde{U}$ and $U/D$ are relevant in determining the scaling curve, but occur solely in the combination $\Delta_0D^r/U (=\tilde{U}^{r-1}(D/U)^r)$.  This too is readily understood, since for $|\epsilon_i|=U/2\gg D$ the impurity-host coupling is controlled by the hybridization $\Delta_{\subi}(D)=\Delta_0D^r$ which, togther with $U$, sets the natural energy scales upon whose ratio the phase boundary thus depends (NRG calculations for the regime $U/D \gg 1$ have not to our knowledge hitherto been reported, and will be given in a subsequent publication [31]).  We emphasize however that the limit $U/D \ra \infty$ does {\it not} imply the total suppression of charge fluctuations close to the limiting SC/LM phase boundary of Figure 8, since $\Delta_0D^r/U\equiv \Phi(r)$ remains in general {\it finite} at the transition.  Charge fluctuations only become negligible as $r \ra 0$, where $\Phi(r)\sim r$ and the condition equation (6.3) for applicability of the Schrieffer-Wolff transformation holds.  For this reason the phase boundary of the Anderson model, even as $U/D \ra \infty$, differs from that for the Kondo model save for $r \ra 0$.  Nonetheless, as one expects physically for $U/D \gg 1$, the critical $\Delta_0D^r/U$ remains closer to its `Kondo asymptote' of $\pi r/8$ over a wider $r$-range than arises for $U/D \ll 1$ (Figure 5).  This is seen in Figure 8 where the Kondo asymptote is reached in practice for $r \lesssim 0.1$, and departure from it is modest for all $r<\frac{1}{2}$. 

\begin{figure} 
\begin{center} 
\epsfig{file =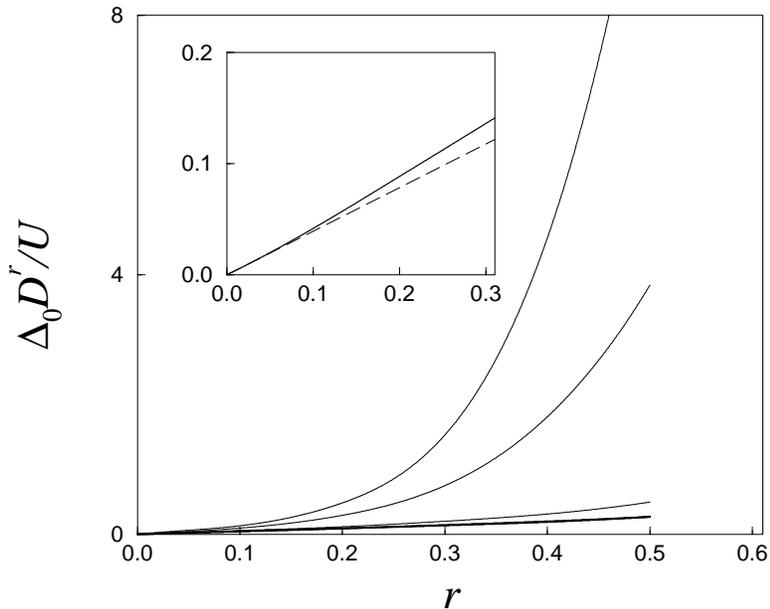,width=10cm} 
\caption{LMA phase boundaries $(\Delta_0D^r/U)_{\cc}$ versus $r$, for (top to bottom) $U/D$ = $10^{-3}$, $10^{-2}$, 1, 10 and 100; for $U/D \gg 1$ common scaling ocurs (bold line), as discussed in text.  Inset: limiting $U/D \gg 1$ phase boundary on expanded scale, compared to $r \ra 0$ Kondo asymptote (dashed line) of $\pi r/8$} 
\end {center} 
\end{figure} 

\subsection{Comparison to NRG}
We now compare quantitatively the LMA phase boundary with NRG results previously reported for the regime $U/D \ll 1$ by Bulla, Pruschke and Hewson (BPH) [28] and Gonzalez-Buxton and Ingersent (G-BI)[21].  BPH have obtained the phase boundary for $U/D = 0.001$, and G-BI for $U/D= 0.02$ and $0.2$.  These $U/D$'s are sufficiently small that one is in the universal scaling regime where (see Figure 7) the critical $\Delta_0U^r/U$ is independent of $U/D$; as noted above, this is demonstrated explicitly in Figure 20 of [21].  The NRG results for $(\Delta_0U^r/U)_{\cc}$ versus $r$ are shown in Figure 9, together with the corresponding LMA result (for the wide band limit $U/D\ra0$, as given in Figure 5); the inset shows the data for $r<0.3$ on an expanded scale, and includes the $r \ra 0$ Kondo asymptote of $\pi r/8$.  It is evident that, except for the important regime $r \ra 0$, the two sets of NRG data do not coincide, although they should: with increasing $r$ the BPH results lie increasingly above those of G-BI.  We believe however that the former results progressively overestimate the phase boundary with increasing $r$, support for which will be given in a subsequent paper [31].

From Figure 9 it is seen that the LMA phase boundary is in excellent agreement with the G-BI results for $r \lesssim 0.3$, where the NRG points essentially lie on the LMA curve.  Further, as discussed in section 6.1, the latter coincides in practice with the Kondo asymptote of $\pi r/8$ for $r \lesssim 0.02$; and this concurs also with the two lowest $r>0$ points from BPH, $r = 0.01$ and $0.02$ (see Figure 9 inset).

\begin{figure}
\begin{center}
\epsfig{file =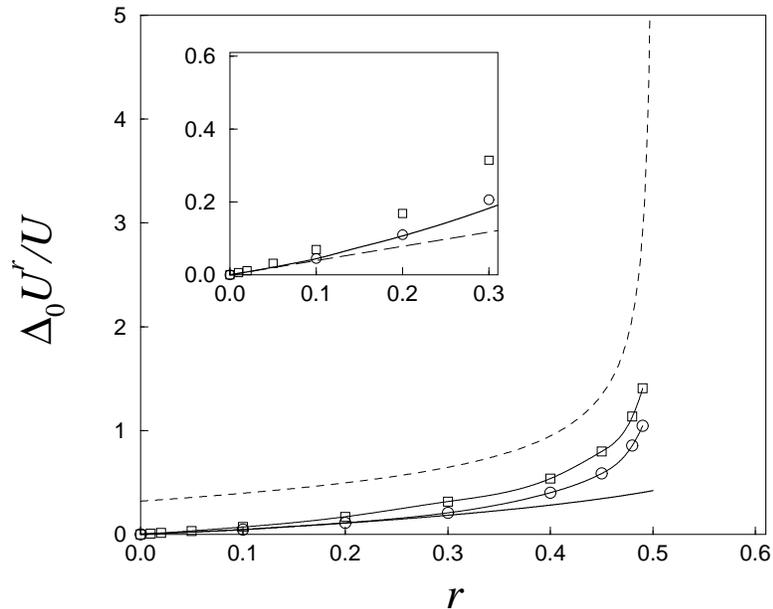,width=10cm} 
\caption{For $U/D \ll 1$, comparison of LMA phase boundary $(\Delta_0U^r/U)_{\cc}$ versus $r$ (solid line) with NRG results: G-BI [21] (circles), BPH [28] (squares); lines connecting NRG points are a guide to the eye only.  The mean-field boundary, diverging as $r \ra \frac{1}{2}-$, is shown for comparison (dashed line).  Inset: results for $r<0.3$ on expanded scale, including the Kondo asymptote $\pi r/8$ (dashed line).}
\end {center}
\end{figure}

For $r\gtrsim0.3$, the NRG phase boundaries $(\Delta_0U^r/U)_{\cc}=\tilde{U}_{\cc}^{r-1}$ increase more rapidly than their LMA counterpart.  That such a deviation should occur is not in itself surprising, since our specific LMA is of course approximate and, although charge fluctuations are obviously included, it seeks primarily to capture the strong coupling physics of the spin-fluctuation regime that is asymptotically dominant for small $r$.  As $r \ra \frac{1}{2}-$, the LMA $(\Delta_0U^r/U)_{\cc}$ tends to a constant value (of $\sim$0.42, see also Figure 5); whereas BPH and G-BI report a divergence in the NRG phase boundary --- as indeed is found at mean-field level where $(\Delta_0U^r/U)_{\cc}$ is known analytically (section 4.1) to diverge as $(1-2r)^{-\frac{1}{2}}$ as $r \ra \frac{1}{2}-$, and which is also included in Figure 9.  For reasons that are evident from the results assembled in Figure 9, we do not however believe that the NRG data themselves warrant such a conclusion: to distinguish numerically between a weakly divergent phase boundary and a finite limit --- the latter of which has been reported in NRG studies of both the symmetric soft-gap Kondo model [27] and the corresponding two-channel Kondo model [21] --- is a delicate matter.  We add however that this remark does not presume a definite answer to the question, for it is possible that the present LMA may not handle adequately the approach to $r=\frac{1}{2}$, which is without doubt a subtle limit worthy of further study.

\seceq
\section{Dynamics: single-particle spectra}
The ability to describe successfully single-particle spectra constitutes a stringent test of any approximate many-body theory.  This is true even for the normal $r=0$ Anderson model, where as mentioned in section 1 current theories have had somewhat limited success; and which is but a limit of the naturally more subtle generic case of $r>0$.  In this section we consider illustrative impurity spectra, $D(\w)$, arising from the present LMA, on all energy scales and for both SC and LM states.  Low-frequency spectral characteristics in the SC phase, and in particular the predicted scaling thereof as the SC/LM phase boundary is approached, will be investigated in section 8.

For a representative $r < \frac{1}{2}$, we begin with an overview of spectral evolution upon decreasing $\tilde{U}=U/\delr$ through the LM phase, into the SC state and towards the weak coupling limit; including comparison of spectra on either side of the SC/LM phase boundary, and a brief discussion of the many-body broadening characteristic of the high-energy Hubbard satellites, which is correctly captured by the present theory.  The weak coupling ($\tilde{U}\ra 0$) behaviour of the spectra is then considered (section 7.1), for both $r <\frac{1}{2}$ where the resultant state is SC (see e.g. Figure 5) and $r>\frac{1}{2}$ where it is LM.  For $r<\frac{1}{2}$ and $r>1$ in particular, where perturbation theory in $U$ about the non-interacting limit is known to be applicable [29], we show that the present theory is perturbatively exact to (and including) second-order in $U$.  Finally, comparison is made (section 7.2) to published NRG results for single-particle spectra in both SC and LM phases [28].

The procedure for determining the impurity Green function $G(\w)$, and hence the single-particle spectrum $D(\w)=-\pi^{-1} \mbox{sgn}(\w)\mbox{Im}G(\w)$, is simply summarized: $G(\w)$ is given by equations (3.2,3), with interaction self-energies $\tilde{\Sigma}_\sigma(\w)$ (to which the symmetry equation (3.5) applies) given by equation (3.8),  the dynamical contribution to which, $\Sigma(\w)$, is given in the present LMA by equation (5.8); and in the SC phase the pinning condition equation (5.9) is enforced as described in section 5.1.  In Figure 10, for $r=0.2$ (and the wide-band limit), we show the resultant dimensionless spectra $D'(\tilde{\w})=\delr D(\w)$ versus $\tilde{\w}=\w/\delr$ upon progressively decreasing $\tilde{U}$ in the LM phase: $\tilde{U} = $100 (a), 25 (b), 20 (c) and 17(d); the critical $\tilde{U}_{\cc}(r) \simeq 15.8$.  Figure 11 continues into the SC phase, with $\tilde{U}= $14 (a), 7 (b) and 2 (c).

\begin{figure}
\begin{center}
\epsfig{file =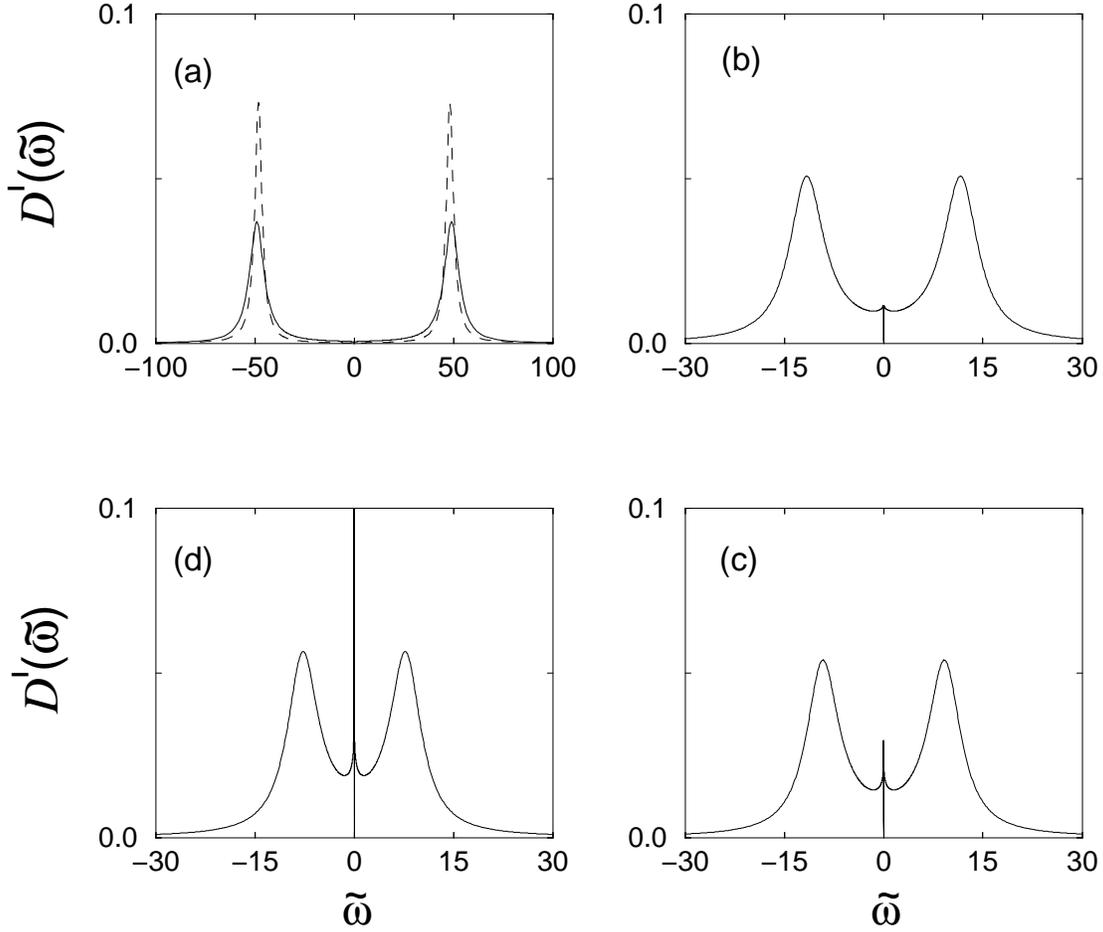,width=15cm} 
\caption{Dimensionless single-particle spectra $D'(\tilde{\w})=\delr D(\w)$ versus $\tilde{\w}=\w/\delr$ for $r=0.2$, with decreasing $\tilde{U}$ in the LM phase: $\tilde{U}= $ 100 (a), 25 (b), 20 (c) and 17 (d); the critical $\tilde{U}_{\cc}(r) \simeq 15.8$.  In (a) the corresponding mean-field spectrum is also shown (dashed line).  Full discussion in text.}
\end {center}
\end{figure}

For $\tilde{U}$ = 100 (Figure 10a), $D'(\tilde{\w})$ is entirely dominated by the Hubbard satellites centred on $\w = \pm U/2$, with no spectral structure apparent in the vicinity of the Fermi level, $\w =0$.  Throughout the LM phase, $D'(\tilde{\w}) \sim |\tilde{\w}|^r$ as $\w \ra 0$ (see equation (5.30b)) and thus vanishes at the Fermi level, as evident in Figure 10.  Upon decreasing $\tilde{U}$ in the LM phase, however, a narrow low energy structure develops in the vicinity of the Fermi level, and becomes increasingly pronounced (Figure 10b-d) as $\tilde{U}$ is decreased towards the LM/SC phase boundary at $\tilde{U}_{\cc}(r)$.  This is a precursor, in the LM phase, of the $|\w|^{-r}$ divergence in $D(\w)$ characteristic of the SC phase as $\w \ra 0$.  The latter in turn is evident in Figure 11 for $\tilde{U}<\tilde{U}_{\cc}$ in the SC phase where, with further decreasing $\tilde{U}$ the Hubbard satellites progressively lose intensity, evolving smoothly  to weak spectral shoulders and naturally vanishing entirely as $\tilde{U} \ra 0$.

\begin{figure}
\begin{center}
\epsfig{file =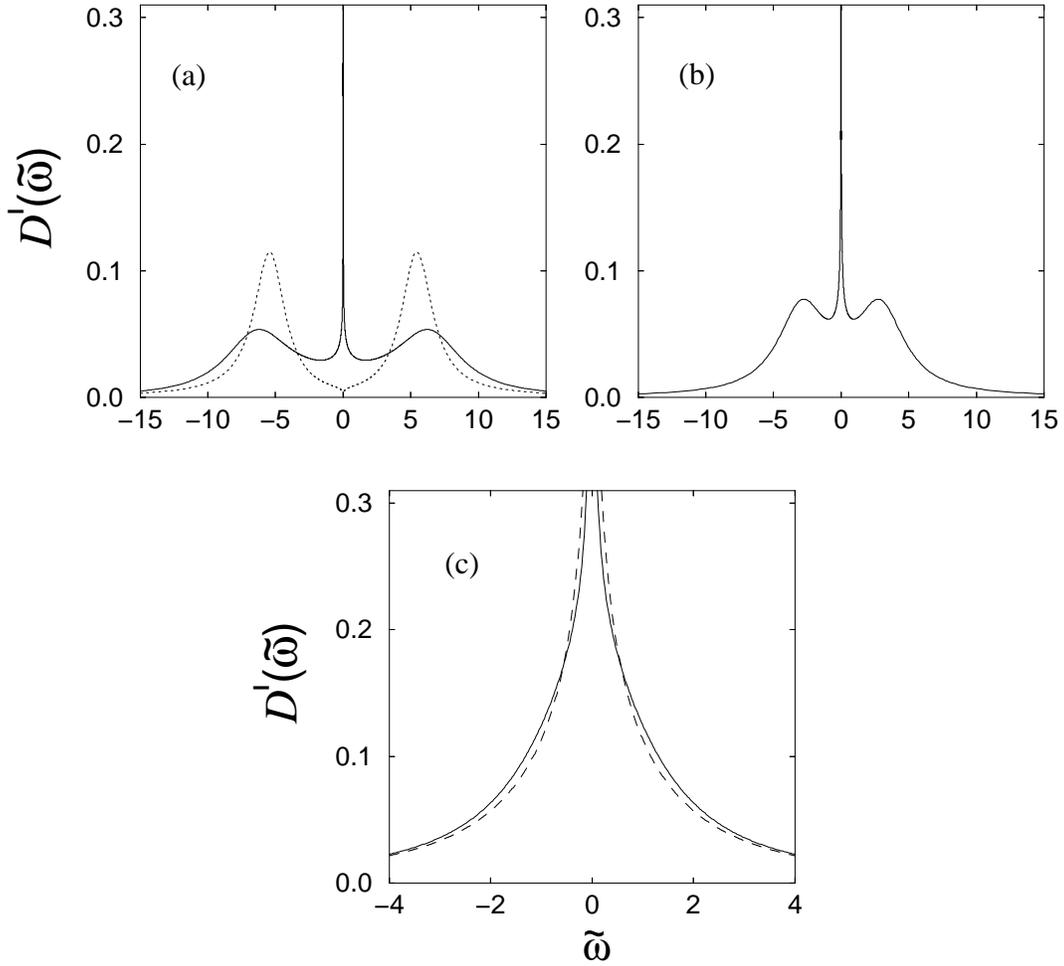,width=15cm} 
\caption{As for Figure 10 but for $\tilde{U}<\tilde{U}_{\cc}$ in the SC phase: $\tilde{U}=$ 14 (a), 7 (b) and 2 (c). In (a) the corresponding mean-field spectrum is also shown (dashed line); and in (c) the spectrum arising from second order perturbation theroy in $U$ (section 7.1) is also shown (dashed line).  Full discussion in text.}
\end {center}
\end{figure}

To illustrate spectral evolution as the LM/SC phase boundary is approached, $D'(\tilde{\w})$ (again for $r=0.2$) is shown in Figure 12 for $\tilde{U} =16.1 $ (LM) and $\tilde{U} = 15.5$ (SC), i.e. for $\tilde{U}/\tilde{U}_{\cc}(r)=1 \pm \delta$ with $\delta \simeq 0.02 \ll 1$; the inset compares the low-frequency behaviour of the spectra for $|\tilde{\w}|< 1 \times 10^{-5}$.  As can be seen, the LM and SC spectra are near coincident on all frequency scales save the lowest (Figure 12 inset); and as $\delta \ra 0$ the LM/SC spectra coincide to arbitrarily low energies, whence the spectra evolve smoothly as the phase boundary is approached.

\begin{figure}
\begin{center}
\epsfig{file=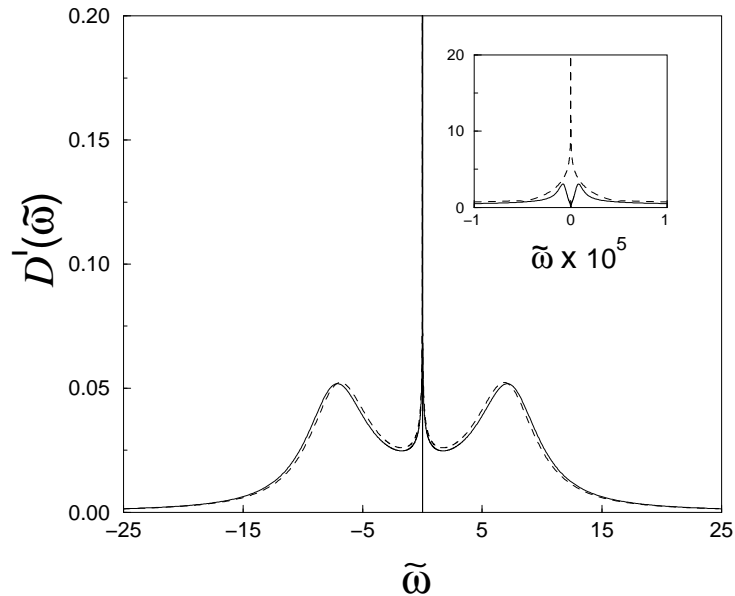,width=10cm} 
\caption{Single particle spectra $D'(\tilde{\w})$ versus $\tilde{\w}$ close to the SC/LM phase boundary.  For $r=0.2$, with $\tilde{U}=16.1$ (LM, solid line) and 15.5 (SC, dashed line); the critical $\tilde{U}_{\cc}(r)=15.8$.  Inset: low-$\tilde{\w}$ behaviour on a much expanded scale.  As $\delta = |1-\tilde{U}/\tilde{U}_{\cc}|\ra 0$, LM and SC spectra coincide to arbitrarily low energies.}
\end {center}
\end{figure}

The behaviour described above is naturally not specific to $r=0.2$, and Figure 13 shows a corresponding spectral series for $r=0.4$.  The principal difference to Figures 10,11 is that, since $\tilde{U}_{\cc}(r=0.4)\simeq 8<\tilde{U}_{\cc}(r=0.2)$ (see Figure 5), the Hubbard satellites are less well developed in the vicinity of the SC/LM phase boundary.

Before proceeding we comment on the strong coupling behaviour of the Hubbard satellite bands in $D(\w)$.  In Figure 10a we superimpose the corresponding UHF spectrum $D_0(\w)=\frac{1}{2}[D^0_\up(\w)+D^0_\down(\w)]$ with $D^0_\sigma(\w)$ given by equation (4.3).  This contrasts strongly with the full LMA spectrum $D(\w)$: the width of the LMA Hubbard satellites is essentially doubled, and their peak heights halved, compared to the mean-field result.  This additional many-body spectral broadening is well known for the normal $r=0$ Anderson model (see e.g. [6,30]).  Its physical origins reflect {\it both} processes illustrated in Figure 2 whereby, having added a $\sigma$-spin electron to a $-\sigma$-spin occupied impurity, either the added $\sigma$-spin or the $-\sigma$-spin already present may hop off the site (Figure 2a and b respectively).  The former alone (`elastic scattering') is captured at UHF level; whereas the latter, involving correlated electron motion is also captured with the present LMA and doubles the rate of electron loss from the impurity site, thus doubling the width of the Hubbard satellites (with concomitant halving of their peak intensity).

\begin{figure}
\begin{center}
\epsfig{file =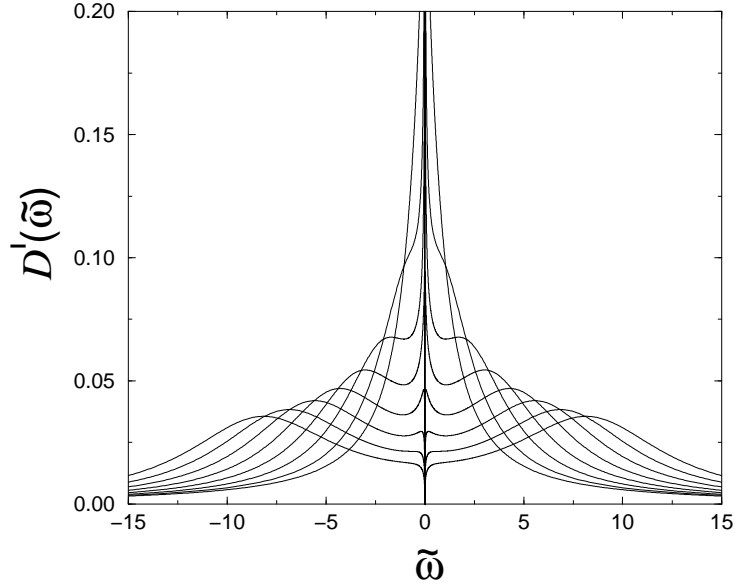,width=10cm} 
\caption{Spectral series $D'(\tilde{\w}) = \delr D(\w)$ versus $\tilde{\w}$ for $r=0.4$ with (top to bottom) $\tilde{U} =$ 3, 6 (SC), 9, 12, 15, 18, 21 and 24 (LM); $\tilde{U}_{\cc}(r)\simeq 8.0$. }
\end {center}
\end{figure}

The additional many-body broadening clearly occurs only if the impurity  level lies within the band, $U/2<D$ (for $U/2>D$ the Hubbard satellites are essentially unbroadened poles).  And its formal origins reside in equation (5.27) for $\Sigma^{\subi}_\up(\w)$ appropriate to the LM phase: as $\tilde{U} \ra \infty$, the poleweight $Q$ (and local moment $|\mu_0|$) naturally saturate to unity, the second term in equation (5.27) is of negligable intensity and $\Sigma^{\subi}_\up(\w)\sim \pi U^2 D^0_\down(\w)\theta(-\w)$; using equation (4.3) for $D^0_\down(\w)$ thus yields $\Sigma^{\subi}_\up(\w)\sim\Delta_{\subi}(\w)$ for frequencies $\w \sim -\frac{U}{2}$ appropriate to the lower Hubbard band (LHB).  $G_\up(\w)$ follows from equation (3.3) with self-energy $\tilde{\Sigma}_{\up}(\w)\stackrel {U \ra \infty}{\sim}-\frac{1}{2}U+\Sigma_\up(\w)$; and for $\w \sim -\frac{1}{2}U$ the impurity Green function itself is given as $U \ra \infty$ by $G(\w) \sim \frac{1}{2}G_\up(\w)$ (as follows from equation (3.2) noting that $D_\down(\w)$ is centred on $\w \sim +\frac{1}{2}U$). Hence, for frequencies $\w \sim -U/2$ in the LHB,
\be
G(\w)=\stackrel{U \ra \infty}{\sim}\frac{\frac{1}{2}}{\left(\w+\frac{U}{2}\right)-2\mbox{i}\Delta_{\subi}(\w)}\ \ \ \ \ :\w \sim -\mbox{$\frac{U}{2}$}
\ee
(where we have neglected $\Delta_{\subr}(\w)$ and $\Sigma^{\subr}_\up(\w)$ since these merely induce a small shift in the satellite positions that vanishes as $U \ra \infty$); the analogous result for the upper Hubbard band follows from particle-hole symmetry, $G(-\w)=-G(\w)$.  By contrast, the corresponding mean-field result $G_0(\w)\sim \frac{1}{2}{\cal G}_\up(\w)$ produces
\be
G_0(\w)\stackrel{U \ra \infty}{\sim}\frac{\frac{1}{2}}{\left(\w + \frac{U}{2} \right) - \mbox{i}\Delta_{\subi}(\w)} \ \ \ \ \ :\w \sim -\mbox{$\frac{U}{2}$}
.\ee
The additional spectral broadening, and consequent halving of the satellite peak intensities, is evident from equation (7.1) in comparison to its mean-field counterpart equation (7.2); and equation (7.1) is found to provide a numerically accurate description of the Hubbard satellites in strong coupling, for any $r$ in the LM phase.

The above qualitative behaviour is not however confined exclusively to strong coupling, but persists in practice throughout the LM regime and (for $r<\frac{1}{2}$) into the SC state.  This is seen in Figure 11a for $r=0.2$ and $\tilde{U}=14$ in the SC phase: the Hubbard satellites in $D(\w)$ remain centred on $\w \sim \pm U/2$ and, in comparison to the mean-field $D_0(\w)$ (superimposed on the Figure), the additional many-body broadening and halving of the satellite peak intensities remains clearly evident.  We add further that as $r \ra 0$, where $\tilde{U}_{\cc}(r) \sim 1/r$ (see equation (6.11)), the SC phase persists to increasingly large $\tilde{U}$; and equation (7.1) can again be shown to hold asymptotically for the SC state, starting from equation (5.12) for the self-energy appropriate to the SC phase and employing a directly analogous argument to that given in [30] for the $r=0$ Anderson model.

\subsection{Weak coupling}
We consider now the behaviour of the spectra in weak coupling $\tilde{U}\ra 0$, focusing separately on the $r$-regimes $r<\frac{1}{2}$, $r>1$ and $\frac{1}{2}<r<1$, and recalling that for $r<\frac{1}{2}$ and $r>1$ conventional perturbation theory in $U$ about the non-interacting limit is known to be applicable [29].

\tolerance=20
We begin with $r<\frac{1}{2}$, where as $\tilde{U} \ra 0$ the ground state is SC (see e.g. Figure 5).  Upon decreasing $\tilde{U}$ in the SC phase, the local moment $|\mu|$ determined self-consistently from equation (5.9) (see section 5.1) progressively diminishes and vanishes at a $\tilde{U}_0\equiv \tilde{U}_0(r)$ ($<\tilde{U}^0_{\cc}(r)$ (see section 4.1), as found in our previous study of the $r=0$ Anderson model [30]).  For $\tilde{U}<\tilde{U}_0$, $|\mu|=0$ is the sole solution; and since the mean-field propagators ${\cal G}_\sigma(\w)$ (equation (4.2)) depend on $U$ and the spin $\sigma$ solely in the combination $\sigma U |\mu|/2$, both they and the polarization bubbles $^0\!\Pi(\omega)$ (equation (5.3)) are independent of both $\tilde{U}$ and the spin indices, and are given by the non-interacting limit result.  In consequence the interaction self-energies $\Sigma_\sigma(\w)$ (equation (5.8)) are likewise $\sigma$-independent, and hence coincide with the single self-energy $\Sigma(\w)$ defined by equation (2.12) (as follows using equations (3.2,3,8)).  From equation (5.8) for $\Sigma_\up(\w)\equiv \Sigma(\w)$, the $\tilde{U}$-dependence of the self-energy thus arises from the explicit $U^2$ prefactor thereto, together with the $U$-dependence of $(\Pi^{+-}(\w)\equiv)\ \Pi(\w)=^0\!\Pi(\omega)/(1-U^0\!\Pi(\omega))$ entering the self-energy kernel.  But for $\tilde{U}<\tilde{U}_0$, $\Pi(\w)$ may be expanded perturbatively in $U$; and its leading term, $^0\!\Pi(\omega)$, when used in equation (5.8) for $\Sigma(\w)$, recovers precisely the result of conventional second-order perturbation theory (SOPT) in $U$ [29].

Thus, while the primary emphasis of the present LMA is naturally on the strong coupling behaviour dominated by the low-energy spin-flip dynamics, the resultant theory for $r<\frac{1}{2}$ is also perturbatively exact to/including second order in $U$ as $U \ra 0$.  This is seen clearly in Figure 11c for $r=0.2$ and $\tilde{U}=2$, where the full LMA spectrum is compared to its SOPT counterpart.  We also add that it is straightforward to show that the stability condition equation (5.21) required for the SC state is always satisfied therein; and in particular that $\mbox{Re}\Pi^{+-}(\w=0)$ evolves smoothly as $\tilde{U}$ passes through $\tilde{U}_0(r)$, and is both positive definite and finite for all $\tilde{U}<\tilde{U}_{\cc}(r)$ in the SC phase.  In consequence, the LMA spectrum $D(\w)$ likewise evolves smoothly upon decreasing $\tilde{U}$ in the SC phase.

We now turn to $r>1$ where the ground state is LM for {\it all} $\tilde{U}$, including the non-interacting limit $\tilde{U}=0$ (see section 2.1 and [29]).  Here too the present LMA recovers asymptotically the results of conventional SOPT as $\tilde{U} \ra 0$; as seen clearly in Figure 14 for $r=1.5$ with $\tilde{U}=7.5 \times 10^{-4}$ and $U/D=\frac{1}{2}$, where the LMA and SOPT spectra are compared. The spectra are shown on a logarithmic scale to bring out clearly the low-$\w$ behaviour, which as $U \ra 0$ is given by the SOPT result equation (5.31), viz $(\delr D(\w)=)\ D'(\tilde{\w})\sim[12/\pi(\tilde{U}q)^2]|\tilde{\w}|^r$; to which end $[\pi(\tilde{U}q)^2/12]D'(\tilde{\w})$ is plotted.  As seen from Figure 14, the difference between the LMA and SOPT spectra for the chosen $\tilde{U}$ is barely perceptible, and becomes even less so with further decreasing $U$.

\begin{figure}
\begin{center}
\epsfig{file =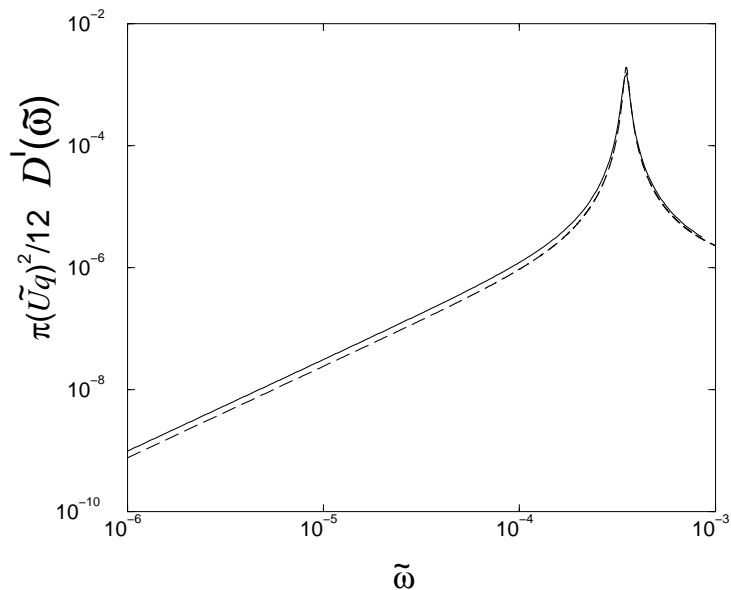,width=10cm} 
\caption{Comparison between LMA (solid line) and SOPT (dashed) spectra, for $r=1.5$ with $\tilde{U}=7.5 \times 10^{-4}$ and $U/D = \frac{1}{2}$; $[\pi(\tilde{U}q)^2/12]D'(\tilde{\w})$ versus $\tilde{\w}$ is shown on a logarithmic scale.  As $\tilde{U} \ra 0$, the LMA recovers SOPT precisely.  Full details in text. }
\end {center}
\end{figure}

In contrast to the $\tilde{U} \ra 0$ behaviour for $r <\frac{1}{2}$, however, this result is rather remarkable.  For $r<\frac{1}{2}$, with a SC ground state as $\tilde{U} \ra 0$, the self-energies $\Sigma_\sigma(\w)$ coincide precisely with each other and with the single self-energy $\Sigma(\w)$, as shown above: all self-energies are equivalent.  But this is not the case for $r>1$.  Here, even as $\tilde{U}\ra 0$, the local moment $|\mu_0|$ remains finite (see equation (4.13)), $\Pi^{+-}(\w)$ is not therefore expandable perturbatively in $U$, and the self-energies $\Sigma_\up(\w)$ and $\Sigma_\down(\w)$ do {\it not} coincide.  Nonetheless, from a knowledge solely of $\tilde{\Sigma}_\up(\w)=-U|\mu_0|/2+\Sigma_\up(\w)$, the single self-energy $\Sigma(\w)$ may still be obtained directly from equation (3.7) which is quite general; as now considered for arbitrary interaction strength, focusing on the salient low-frequency behaviour.  The $\w \ra 0$ behaviour of $\Sigma^{\subi}_\up(\w)$ is given for the LM phase by equation (5.29), and for $r>1$ it is straightforward to show that the corresponding real part has the low-$\w$ behaviour $\Sigma^{\subr}_\up(\w)-\Sigma^{\subr}_\up(0)\sim -\gamma \w$ with $\gamma \equiv \gamma(U)>0$ given by equation (5.18b).  Using this in equation (3.7) leads to the following low-$\w$ behaviour of $\Sigma(\w)=\Sigma^{\subr}(\w)-\mbox{isgn}(\w)\Sigma^{\subi}(\w)$ for $r>1$,
\alpheqn
\be
\Sigma^{\subi}(\w)\stackrel{\w \ra 0}{\sim}\left(\tilde{\Sigma}^{\subr}_\up(0)\right)^2\left[\frac{\pi q}{1+q\gamma(U)}\delta(\w)+\frac{3q^2}{\left(1+q\gamma(U)\right)^2}\frac{\Delta_{\subi}(\w)}{\w^2}\right]
\ee 
\be
\Sigma^{\subr}(\w)\stackrel{\w \ra 0}{\sim}\left(\tilde{\Sigma}^{\subr}_\up(0)\right)^2\frac{q}{1+q\gamma(U)}\ \mbox{P}\left(\frac{1}{\w}\right)
\ee
\reseteqn
(with corrections $\mbox{O}(|\w|^{r-2};|\w|)$ in the latter case); where $q=\left[1-(\partial\Delta_{\subr}(\w)/\partial \w)_{\w=0}\right]^{-1}$ given by equation (2.10b) is the poleweight in the non-interacting single-particle spectrum.

Equation (7.3) is quite general.  But as $U \ra 0$, $\gamma(U)$ may be shown to vanish and $\tilde{\Sigma}^{\subr}_\up(\w =0)$ is dominated by the Fock contribution of $-\frac{1}{2}U|\mu_0|$, with $|\mu_0| \ra q$ as $U \ra 0$ (see equation (4.13)).  Hence, for $U \ra 0$, equation (7.3a) for example reduces to
\be
\Sigma^{\subi}(\w) \stackrel{\stackrel{U\ra 0}{\w \ra 0}}{\sim}\frac{U^2 q^2}{4}\left[\pi q \delta(\w)+3\Delta_0 q^2 |\w|^{r-2}\right].
\ee
This is precisely the result obtained from conventional SOPT for $r>1$; see equations (5.10,11) of reference [29].  And we note that its correct recovery reflects the fact that the prefactor to the $\w$-dependence of $\Sigma^{\subi/\subr}(\w)$ in equation (7.3), viz $\left[\tilde{\Sigma}^{\subr}_\up(0)\right]^2 \sim \left[\frac{1}{2}U|\mu_0|\right]^2$ as $U \ra 0$, is $\mbox{O}(U^2)$ as $U \ra 0$.

The latter remark also sheds some light on the inapplicability of conventional perturbation theory in $U$ for $\frac{1}{2}<r<1$, since within SOPT the prefactor to the $\w$-dependence of $\Sigma^{\subi/\subr}(\w)$ is $\mbox{O}(U^2)$ by construction.  The low-$\w$ behaviour of $\Sigma^{\subi/\subr}(\w)$ in the LM phase for $r<1$ has been considered in section 5.2 and is given generally by equation (5.33), exhibiting a characteristic $|\w|^{-r}$ divergence as $\w \ra 0$ with a $U$-dependent prefactor that is again $\left[\tilde{\Sigma}^{\subr}_\up(0)\right]^2$; and for $\frac{1}{2}<r<1$ the LM phase occurs for all $U>0$ (Figures 5,7,8), hence $U \ra 0+$ can be considered.  As for $r>1$, $\left[\tilde{\Sigma}^{\subr}_\up(0)\right]^2 \sim \left[\frac{1}{2}U|\mu_0|\right]^2$ again as $U \ra 0$; but for $\frac{1}{2}<r<1$ the local moment $|\mu_0|$ is given by equation (4.12) and itself vanishes as $U \ra 0$, whence $\left[\tilde{\Sigma}^{\subr}_\up(0)\right]^2 \sim U^{\frac{2r}{2r-1}}$ as $U \ra 0$.  For $\frac{1}{2}<r<1$ the prefactor to the $\w$-dependence of $\Sigma^{\subi/\subr}(\w)$ thus vanishes as $U \ra 0$, but with an exponent $\epsilon = 2r/(2r-1)$ that is in general non-integral and strictly greater than 2.  Such behaviour cannot by construction be captured by SOPT (or indeed by conventional perturbation theory to any finite order in $U$); an inability that is not unexpected when one recalls that the ground state of the strict non-interacting limit $U = 0$, is SC for all $r <1$ [29].

We emphasize, however, that while the behaviour in the LM phase of the {\it single} self-energy $\Sigma(\w)$ --- obtained as a byproduct of the LMA via equation (3.7) --- is both singular and dependent upon whether $r\gtrless 1$ (cf equations (7.3, 5.33)), the same does not hold for either the self-energies $\tilde{\Sigma}_\sigma(\w)$ central to the LMA (see section 5.2) or the single-particle spectra; in particular, the low-$\w$ behaviour of $D(\w)$ for all $U>0$ in the LM phase is $D(\w)\sim |\w|^r$ (equation (5.30b)) for {\it any} $r>0$.

\subsection{Comparison to NRG}
We now compare predicted LMA spectra with the NRG results of Bulla, Pruschke and Hewson [28], obtained for a fixed $U/D = 10^{-3}$ using a discretization parameter $\Lambda = 2$ ($\Lambda \ra 1$ recovers the continuum limit) and with $\sim800$ states retained at each NRG iteration.  The two sets of spectra are shown in Figure 15 as a function of $\w/D$, on a logarithmic scale to show clearly the low-$\w$ behaviour; for (i) $r=0.25$ and $\Delta_0 = 0.02$ (SC state, $\tilde{U}\simeq 0.18$); (ii) $r=0.25$ and $\Delta_0 = 0.0002$ (LM state, $\tilde{U}\simeq 85$); (iii) $r=0.75$ and $\Delta_0=0.02$ (LM state, $\tilde{U}=6.25 \times 10^3$).
\begin{figure}
\begin{center}
\epsfig{file =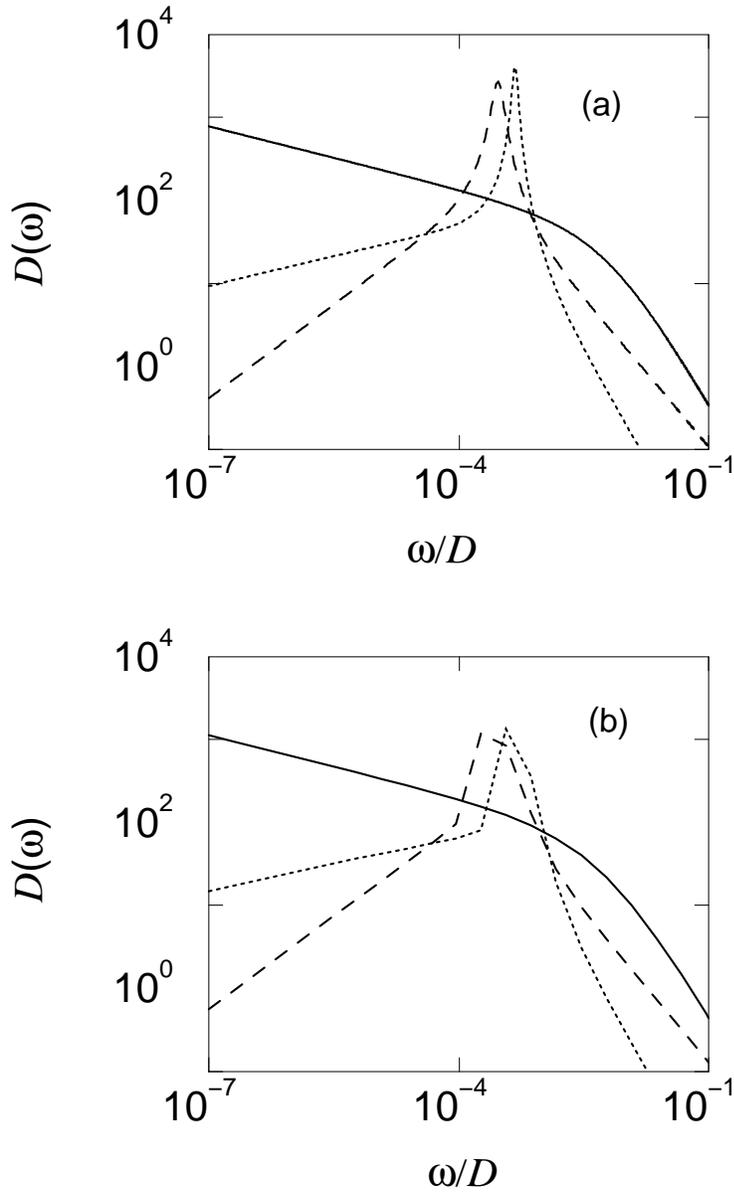,width=10cm} 
\caption{Comparison between LMA spectra (Figure 15a) and NRG spectra [28] (Figure 15b) for fixed $U/D = 10^{-3}$. (i) $r=0.25, \Delta_0=0.02$ (SC, solid line); (ii) $r=0.25, \Delta_0=0.0002$ (LM, dotted); (iii) $r=0.75$, $\Delta_0=0.02$ (LM, dashed).  Spectra are shown on a logarithmic scale, as a function of $\w/D$}
\end {center}
\end{figure}
The characteristic $\w \ra 0$ behaviour of the two phases is clearly seen, $D(\w)\sim |\w|^r$ (LM) and $\sim|\w|^{-r}$ (SC).  The two LM spectra have pronounced Hubbard satellites, that for $r=0.75$ being centred precisely $\w = U/2$; while in the SC spectrum --- which is a weak coupling example ($\tilde{U}\simeq 0.18$) --- the satellites are no longer a distinct feature and have been absorbed into the band. 

The agreement between the LMA and NRG results is self-evident.  For the SC example in particular, the agreement is essentially perfect for all $\w$ when a small, controlled degree of spectral broadening is used to smooth the (necessarily discrete) NRG data [35]; and the $\w \ra 0$ behaviour of the NRG data is readily shown to coincide precisely with that of the non-interacting limit (equation (2.11)), as we have shown is required for the SC state (see section 2.2 and [29]).

Finally, we note that the present LMA includes in $\Sigma_\sigma(\w)$ the sum of all particle-hole interactions in the transverse spin channel (Figure 3a) --- that captures the low-energy spin-flip physics (section 5) --- and one can of course additionally include repeated particle-particle and `bubble' interactions (see figure 9 of [30]).  However as in reference [30] for the normal $r=0$ Anderson model, these have a very minor effect on predicted spectra and are not therefore considered explicitly in the present work.

\seceq
\section{Spectra: pinning and scaling}
We turn now to a rather subtle, and superficially hidden, aspect of the problem: scaling of single-particle spectra in the SC phase ($r<\frac{1}{2}$) as the SC/LM phase boundary is approached.  Such behaviour  is of course well known for the normal $r=0$ Anderson model (see e.g. [2,4,15]) where with increasing $\tilde{U}$ the width of the Kondo resonance in $D(\w)$ becomes exponentially small, reflecting the exponential diminution of the Kondo scale $\w_K$ (see equation (6.13)); and the Kondo resonance becomes a universal function of $\w/\w_K$.

The $r=0$ universal scaling curve has three essential characteristics. (i) It is pinned at the Fermi level: $\pi\Delta_0D(\w=0)=1$ for all $\tilde{U}$; (ii) with characteristic low-frequency Fermi liquid behaviour: $D(\w)-D(0)\sim [\w/\w_K]^2$ for $\w/\w_K \ll 1$. (iii) On larger $(\w/\w_K)$-scales the spectrum follows a Doniach-\u{S}unji\'{c} law [13-15] indicative of the orthogonality catastrophe, whereby $D(\w)\sim [|\w|/\w_K]^{-\alpha}$ with $\alpha = 1-2[\delta_0/\pi]^2$ and $\delta_0=\pi/2$ the Fermi level phase shift; i.e. $D(w)\sim[|\w|/\w_K]^{-\frac{1}{2}}$.  This has been observed in both a QMC study of the $r=0$ symmetric spin-$\frac{1}{2}$ Anderson model [15] and an NRG study of $D(\w)$ for $\w<0$ in the asymmetric case [14]; in practical terms, Doniach-\u{S}unji\'{c} (DS) behaviour in the scaled $D(\w)$ sets in for $\w/\w_K \gtrsim 1$ [14,15].

The question arises as to the whether the above behaviour is specific to the $r=0$ Anderson model, a conventional Fermi liquid; or whether it is but a particular example of behaviour generic to the SC phase for any $r<\frac{1}{2}$.  We show it to be the latter, and to be directly and generally apparent not in $D(\w)$ itself but in the modified spectral function $A(\w)=|\w|^rD(\w)$.  That this is so is natural, since (a) it is $A(\w)$ that is pinned at the Fermi level $\w =0$ for any $r \geq 0$ where a SC state obtains (section 2.2 and [29]); and (b) in $A(\w)$ the unrenormalized $|\w|^{-r}$ divergence in $D(\w)$ has been removed, thus `exposing' directly the low-frequency many-body renormalization characteristic of the Kondo effect, and hence the Kondo scale as shown below.

Specifically, we focus on ${\cal F}(\w)$ defined by
\be
{\cal F}(\w)=\pi\Delta_0 \left[1+\mbox{tan$^2(\frac{\pi}{2}r)$}\right]|\w|^r D(\w)
\ee
which, in the SC phase, is pinned at the Fermi level (equation 2.16b): ${\cal F}(\w=0)=1$ for any $r$.  Figure 16 illustrates ${\cal F}(\w)$ versus $\tilde{\w}=\w/\delr$ for $r=0.2$ in the wide-band limit, with $\tilde{U}=5,10$ and 15 ($<\tilde{U}_{\cc}(r)\simeq 15.8)$.  Spectral pinning is evident.  More significantly, so too is a generalized Kondo resonance in ${\cal F}(\w)$ which has (a) cusp behaviour as $\w \ra 0$, and (b) a width (Kondo scale) that narrows progressively upon increasing $\tilde{U}$ towards $\tilde{U}_{\cc}(r)$ and vanishes as $\tilde{U} \ra \tilde{U}_{\cc}(r)-$.  We show below that the latter is the spin-flip scale $\w_{\m}(r)$ discussed in detail in sections 5 and 6.

\begin{figure}
\begin{center}
\epsfig{file =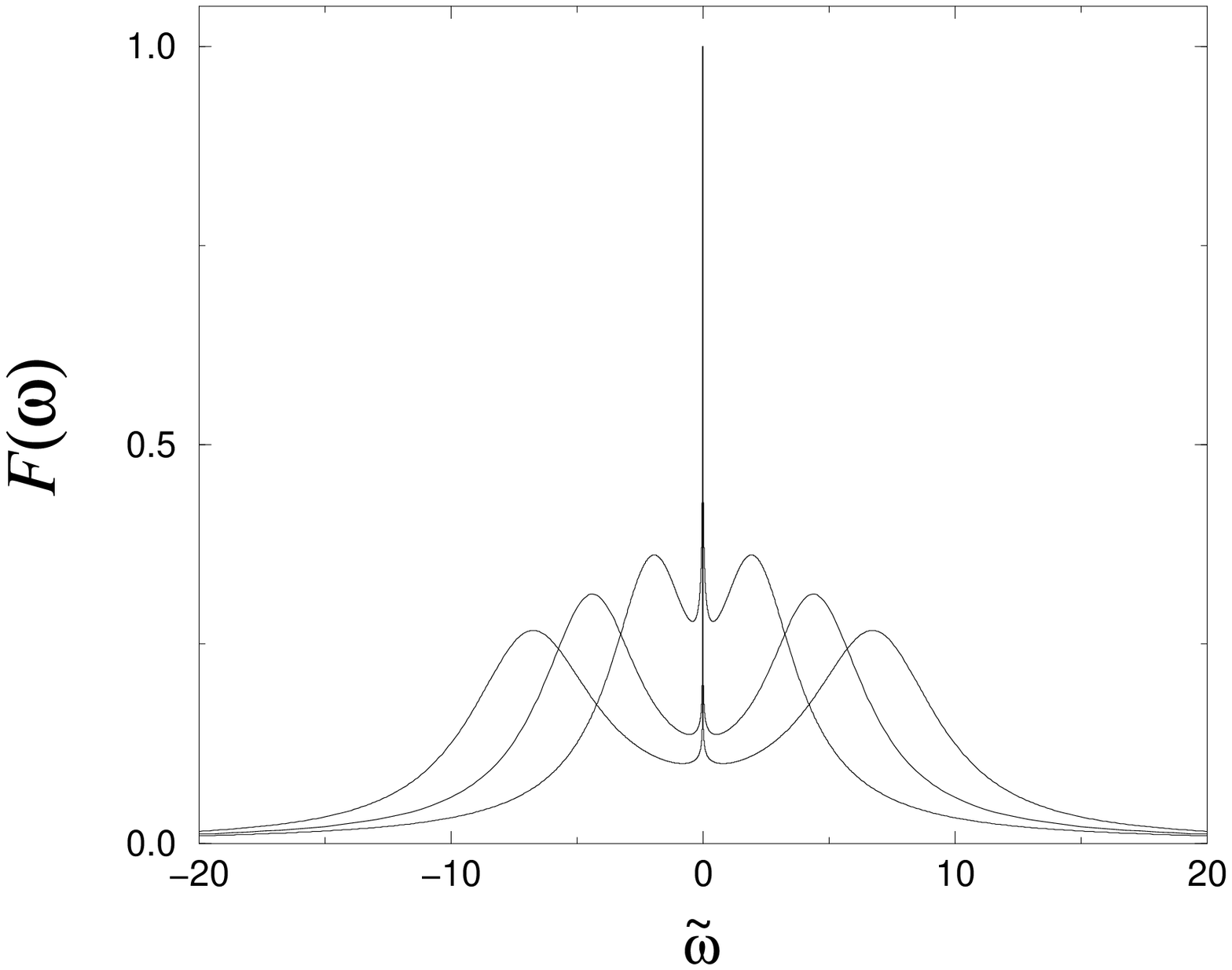,width=10cm} 
\caption{${\cal F}(\w)$ verus $\tilde{\w}=\w/\delr$ in the SC phase; for $r=0.2$ (wide-band limit) with $\tilde{U}=5,10$ and 15.  Note the spectral pinning (${\cal F}(0)=1$) and diminishing width of the Kondo resonance with increasing $\tilde{U}$ towards $\tilde{U}_{\cc}(r)\simeq 15.8$ where $\w_{\m} \ra 0$.}
\end {center}
\end{figure}

First we establish the low-$\w$ behaviour of ${\cal F}(\w)$, and its cusp characteristics.  This follows from $D(\w)=\frac{1}{2}[D_\up(\w)+D_\down(\w)]$ with $D_\sigma(\w)=-\pi^{-1}\mbox{sgn}(\w)\mbox{Im}G_\sigma(\w)$, using equation (3.3) for $G_\sigma(\w)$ together with equations (2.5,9) for $\Delta_{\subi/\subr}(\w)$ and $\tilde{\Sigma}^{\subr}_\sigma(\w) \sim -\gamma \w$ as $\w \ra 0$ (equation (5.19)). (Since $\tilde{\Sigma}^{\subi}_\sigma(\w)$ decays to zero more rapidly than $\tilde{\Sigma}^{\subr}_\sigma(\w)$, see equation (5.17), it is irrelevant to the leading low-$\w$ behaviour.) The result is
\be
{\cal F}(\w) \stackrel {\tilde{\w}\ra 0}{\sim} 1-\mbox{sin($\pi r$)}(q^{-1}+\gamma)|\tilde{\w}|^{1-r} -(q^{-1}+\gamma)^2 \phi(r)|\tilde{\w}|^{2(1-r)}
\ee
with $q^{-1}$ given by equation (2.10b) (here for $r<\frac{1}{2}$); and $\phi(r)=\mbox{cos$^2(\frac{\pi}{2}r)$}[1-4\mbox{sin$^2(\frac{\pi}{2}r)$}]$ such that $\phi(0) = 1$.  For $r>0$, ${\cal F}(\w)$ as $\tilde{\w} \ra 0$ is thus dominated by an $|\tilde{\w}|^{1-r}$ cusp (seen in Figures 16 and 19b); but its prefactor vanishes as $r \ra 0$, and for $r=0$ the parabolic Fermi liquid behaviour ${\cal F}(\w)-1\sim\tilde{\w}^2$ characteristic of the normal Anderson model is recovered.

We now determine the behaviour of ${\cal F}(\w)$, equation (8.2), as $\tilde{U}\ra \tilde{U}_{\cc}(r)-$ where the low-energy spin-flip scale $\w_{\m}(r)$ vanishes as described in sections 5.1 and 6.2; and for which $\gamma = -\left( \partial \tilde{\Sigma}^{\subr}_\up(\w)/\partial \w \right)_{\w = 0}$ is thus required.  The full self-energy $\tilde{\Sigma}^{\subr}_\up(\w) = -\frac{1}{2}U|\mu|+\Sigma^{\subr}_\up(\w)$ is given generally in the SC phase by $\tilde{\Sigma}^{\subr}_\up(\w)={\Sigma}^{\subr}_\up(\w)-\Sigma^{\subr}_\up(0)$ (from equation (5.9)); and the low-$\w$ behaviour of $\Sigma^{\subr}_\up(\w)$ may be extracted analytically in strong coupling $\tilde{U}\gg 1$ where, from a trivial extension of the argument given in section 6.1 leading to equation (6.4), $\Sigma^{\subr}_\up(\w)\sim U^2 \mbox{Re}{\cal G}^-_\down(\w+\w_{\m})$.  Using equation (6.8b) (with $x \equiv \frac{1}{2}U$) for the $\w >0$ behaviour of $\mbox{Re}{\cal G}^-_\down(\w)$ then yields
\be
\tilde{\Sigma}^{\subr}_\up(\w)\stackrel {\w \ra 0}{\sim}-\frac{4\Delta_0}{\mbox{sin$(\pi r)$}}\w_{\m}^r\left[\left(1+\frac{\w}{\w_{\m}}\right)^r-1\right]
\ee
(with corrections $\mbox{O}\left[(\w_{\m}/\lambda)(1+\w/\w_{\m})\right]$) where $\lambda = \mbox{min}\left[D,\frac{U}{2}\right]$; and hence for $\w/\w_{\m} \ll 1$ the requisite low-$\w$ behaviour
\be
\tilde{\Sigma}^{\subr}_\up(\w) \stackrel{\w \ra 0}{\sim} -\frac{4r}{\mbox{sin($\pi r$)}}\tilde{\w}_{\m}^{r-1}\w\equiv -\gamma\w.
\ee

From equations (8.2,4) the asymptotic behaviour of ${\cal F}(\w)$ as the SC/LM phase boundary is approached, $\tilde{U}\ra \tilde{U}_{\cc}(r)-$, is thus given by
\be
{\cal F}(\w)\sim 1-4r\left(\frac{|\tilde{\w}|}{\tilde{\w}_{\m}}\right)^{1-r}-\left(\frac{4r}{\mbox{sin($\pi r$)}}\right)^2\phi(r)\left(\frac{|\tilde{\w}|}{\tilde{\w}_{\m}}\right)^{2(1-r)}
\ee
(where the terms involving $q^{-1}$ drop out since $\w_{\m}(r)\ra 0$).  The specific $r$-dependent coefficients here are naturally valid asymptotically as $r \ra 0$, since the above analysis hold strictly as $\tilde{U}_{\cc}(r)\ra \infty$ (see section 6.1 and equation (6.11)); although in practice $\tilde{U}_{\cc}(r) \gg 1$ for all $r<\frac{1}{2}$ (see e.g. Figure 5).  The important point of course is that ${\cal F}(\w)$ exhibits universal scaling; and that, as expected physically, it is indeed the spin-flip or Kondo energy $\w_{\m}(r)$ that sets the scale for such behaviour.  Note relatedly that the physical content of equation (5.20) for the SC phase is the retoration, for sufficiently long times, of the locally broken symmetry inherent to the zeroth order mean field level of description; and that this timescale is $1/\w_{\m}$ (as for the normal Anderson model [30], the result for which is the $r=0$ limit of equation (8.4)).  For the doubly degenerate LM state by contrast there is naturally no such symmetry restoration, reflected in the fact that $\Sigma_\up(\w)$ and $\Sigma_\down(\w)$ do not coincide as $\w \ra 0$ (see equation (5.29)).

The analysis above, while demonstrating universality, is confined to the low frequency behaviour $\w/\w_{\m} \ll 1$.  We now consider the scaling behaviour of ${\cal F}(\w)$ over the entire $\w/\w_{\m}$ range.

\subsection{Scaling of ${\cal F}(\w)$}
We begin with the $r=0$ Anderson model, for which ${\cal F}(\w)\equiv \pi \Delta_0 D(\w)$.  The resultant ${\cal F}(\w)$ versus $\w/\w_{\m}$ obtained from from the LMA is shown in Figure 17, which universal form is reached in practice for $\tilde{U} \gtrsim 5\pi$ [30].  This scaling spectrum was itself obtained in [30], but one important facet of it was not noted (and we are grateful to A E Ruckenstein for drawing it to our attention): the DS tails for $\w/\w_{\m} \gtrsim 1$ are captured by the theory.  This is seen in Figure 17 where an $(\w/\w_{\m})^{-1/2}$ fit is made to the wings of the spectrum.  Detailed comparision between the LMA scaling spectrum and NRG calculations will also be given in a subsequent paper [31]; suffice it here to say that the agreement is rather good.  To our knowledge the LMA is the only theoretical approach that captures simultaneously the characteristic low-$\w$ Fermi liquid behaviour $D(\w)-D(0)\sim [\w/\w_{\m}]^2$, and the DS tails for $\w/\w_{\m} \gtrsim 1$ that are known to arise experimentally (see e.g. [16]) but to be washed out in e.g. slave boson or $1/N$-expansion approaches [10-12].

\begin{figure}
\begin{center}
\epsfig{file =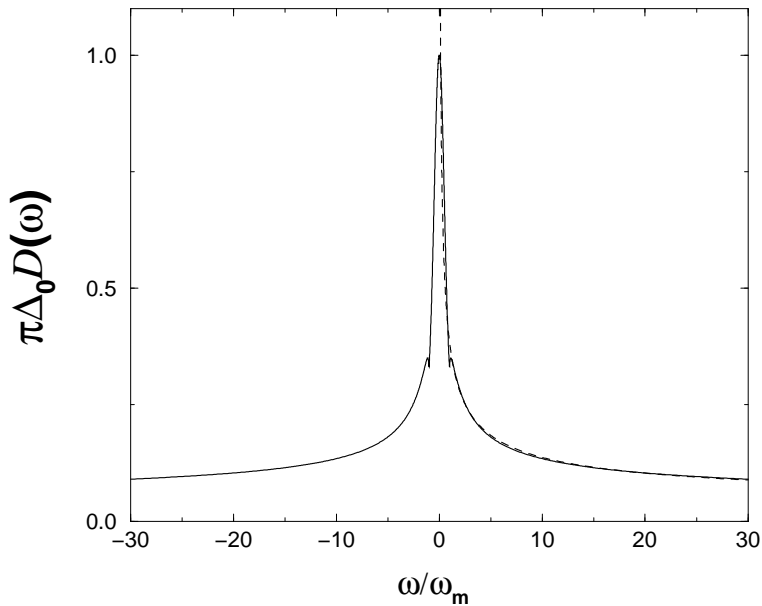,width=10cm} 
\caption{Scaling for normal ($r=0$) AIM: ${\cal F}(\w)\equiv \pi \Delta_0 D(\w)$ versus $\w/\w_{\m}$.  The Doniach-\u{S}unji\'{c} tails are seen from the $(\w/\w_{\m})^{-\frac{1}{2}}$ fit to the wings of the spectrum (dashed line).}
\end {center}
\end{figure}

We also point to the apparent small spectral feature occuring in Figure 17 at $\w/\w_{\m} \sim 1$.  As discussed in [30] this is entirely an artifact of using the specific form equation (5.2) for $\Pi^{+-}(\w)$ in equation (5.8) for the self-energy.  This is not however an integral element of the LMA and (as discussed in [30]) may be circumvented, thereby eliminating the spectral anomaly but otherwise producing no significant effect on either the scaling spectrum (see figure 12 of [30]) or previously deduced asymptotics.  The same feature is naturally present also in the $r>0$ spectra below, and can likewise be removed; but it is a minor effect that we are content to live with in the following.

For $r=0.2$, the resultant ${\cal F}(\w)$ for the wide-band limit is illustrated in Figure 18 for $\tilde{U}=$ 10, 13 and 15.  The inset shows the central portion of the Kondo resonance --- whose half-width is proportional to $\w_{\m}(r)$, as for the $r=0$ model [30] --- versus $\tilde{\w} = \w/\delr$ on an `absolute' scale, to illustrate its rapid narrowing with increasing $\tilde{U}$ and vanishing as $\tilde{U} \ra \tilde{U}_{\cc}(r)- \simeq 15.8$.  The main Figure by contrast shows ${\cal F}(\w)$ versus $\w/\w_{\m}$, from which universality is evident; and note that although $\tilde{U}_{\cc}(r)$ is finite for all $r>0$, the Hubbard satellites are again eliminated from the scaling spectrum since $\w_{\m}(r) \ra 0$ as $\tilde{U}\ra \tilde{U}_{\cc}(r)-$.  We add moreover that while $\tilde{U}_{\cc}(r)$ itself depends on the host bandwidth $D$ (see Figures 7,8), the latter has no detectable influence on the scaling spectrum; as is expected physically, and indeed seen from equations (8.2,5) where the $D$-dependence of the former (contained in $q^{-1}$) is eliminated in equation (8.5) as $\w_{\m}(r) \ra 0$.

\begin{figure}
\begin{center}
\epsfig{file =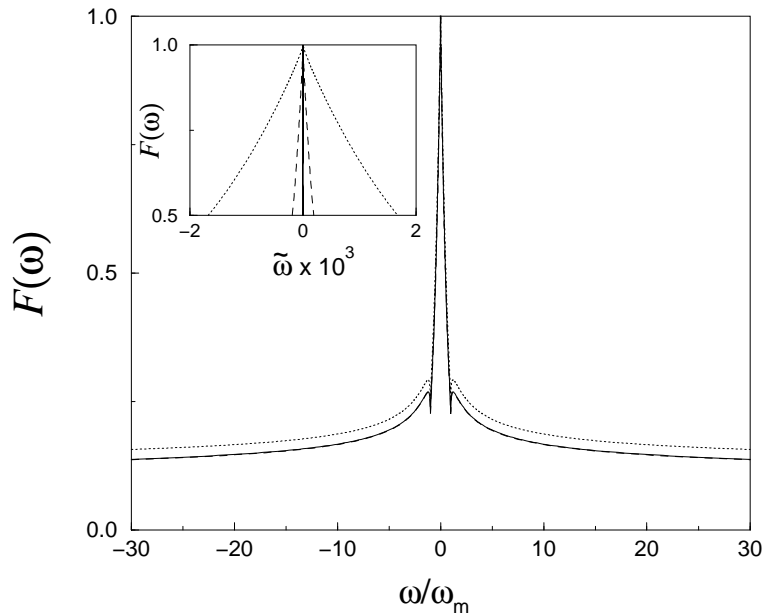,width=10cm} 
\caption{Scaling behaviour of Kondo resonance for $r>0$ SC phase: for $r=0.2$, ${\cal F}(\w)$ versus $\w/\w_{\m}$ for $\tilde{U}$ = 10 (dotted), 13 (dashed) and 15 (solid).  Inset: corresponding ${\cal F}(\w)$ versus $\tilde{\w}=\w/\delr$ on an `absolute' scale, to show narrowing of Kondo resonance upon increasing $\tilde{U}$ towards SC/LM transition at $\tilde{U}_{\cc}(r)\simeq 15.8$ where $\w_{\m}\ra 0$.}
\end {center}
\end{figure}

Finally, Figure 19a compares the universal scaling spectra ${\cal F}(\w)$ versus $\w/\w_{\m}$ for $r=$ 0, 0.2 and 0.4; while Figure 19b shows the data on a reduced scale $\w/\w_{\m} <1$ for five different $r$-values, to illustrate in particular the evolution of the low-$\w$ cusp behaviour (equation (8.5)).  From Figure 19a it is seen that, as for the $r=0$ case, DS tails again arise in the scaling spectrum for $\w/\w_{\m} \gtrsim 1$.  Numerical analysis shows these to have the form ${\cal F}(\w)\sim [\w/\w_{\m}]^{-\nu}$ with exponent $\nu(r)=\frac{1}{2}-r$, thus `flattening out' with increasing $r$ as evident in Figure 19a.  Note moreover from equation (8.1) that while $D(\w)$ itself is not a universal function of $\w/\w_{\m}$, $\w_{\m}^r D(\w) \propto (|\w|/\w_{\m})^{-r}{\cal F}(\w)$ does exhibit scaling.  Its DS tail behaviour is thus $\sim (|\w|/\w_{\m})^{-1/2}$ as for the normal $r=0$ model; and we note that this conforms to the DS law exponent of $\alpha = 1-2[\delta_0/\pi]^2$, since for $r>0$ the phase shift $\delta_0$ precisely at the Fermi level is readily shown to be $\pi/2$, as for the $r=0$ case.

\begin{figure}
\begin{center}
\epsfig{file =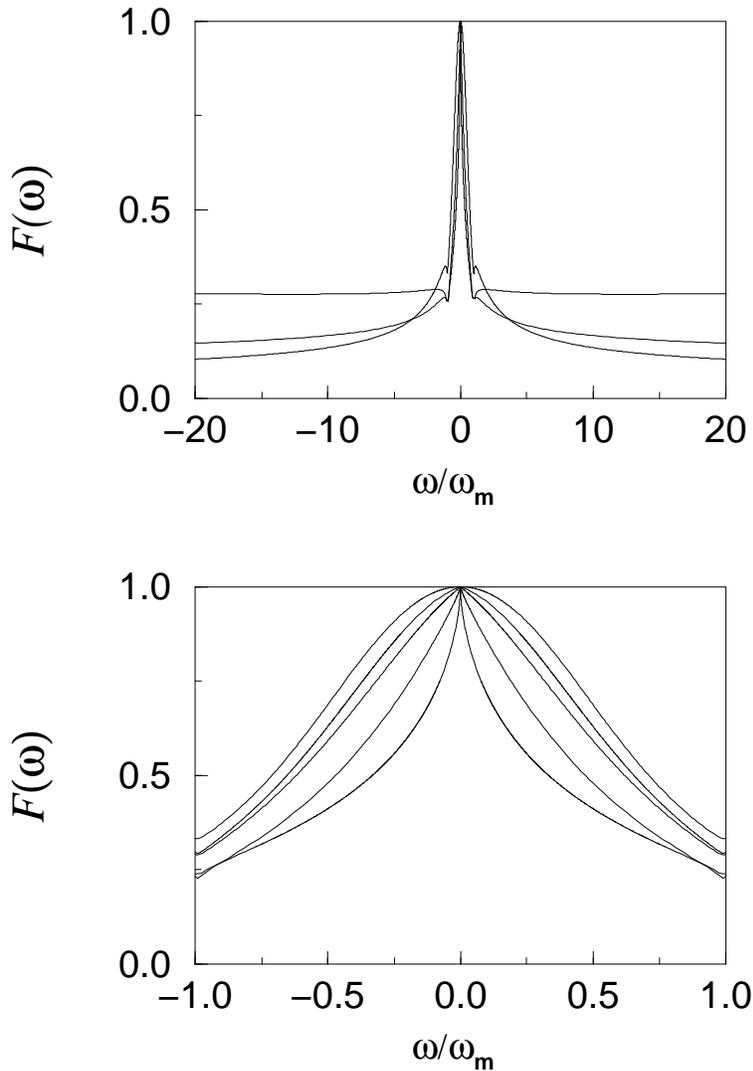,width=10cm} 
\caption{Universal scaling spectra ${\cal F}(\w)$ versus $\w/\w_{\m}$.  (a) For $r=$ 0 0.2 and 0.4.  DS tails arise in all cases, and `flatten out' with increasing $r$; details in text.  (b) For $\w/\w_{\m}<1$ and $r=$ 0, 0.05, 0.1, 0.2 and 0.4 (from outside to inside), to illustrate evolution of low-$\w$ `cusp' behaviour.}
\end {center}
\end{figure}

For the $r \geq 0$ SC phase generally, the local moment approach thus predicts universal scaling of the modified spectral function ${\cal F}(\w)$ (equation (8.1)) as the SC/LM phase boundary is approached; with characteristic Fermi level pinning, low-$\w$ cusp behaviour and DS tails in ${\cal F}(\w)$.  In a subsequent paper [31] we shall see that these predictions are borne out by NRG calculations, to which detailed comparison will be made.      

\section{Summary}

We have developed in this paper a many-body local moment approach to
the symmetric soft-gap Anderson impurity model, including the
`normal' ($r=0$) Anderson model as a particular limit [30]. The LMA is
naturally non-perturbative, and both the notion of local moments and
the {\it a priori}  possibility of either a SC or LM state are introduced
explicitly and  {\it self-consistently}  from the outset; as reflected in the 
employment of an underlying two-self-energy description, together with 
self-consistent imposition of the $U$-independent spectral pinning 
condition at the Fermi level that is characteristic of the SC phase [29]. 
  
  The primary emphasis of the LMA is on single-particle dynamics ---  
posing well known and hitherto unsurmounted difficulties for traditional theories --- 
but an integral element of the approach also permits direct 
analysis of the SC/LM transition and associated phase boundaries. The theory 
offers a rather comprehensive description of both SC and LM phases, 
for any $r\geq 0$. The entire range of interaction strengths is also covered, 
including the Kondo/spin-fluctuation physics that dominates the SC phase 
at large interaction strengths, and the `cost-free' spin-flip physics of the LM state;
as well as the weak coupling (small-$U$) behaviour that is not as prosaic as 
naive expectation might suggest, being intrinsically non-perturbative for 
$\frac{1}{2}<r<1$. 

   While the theory leads to very good agreement with extant NRG calculations 
[21,28], a significant number of further predicitions arise from it that can likewise 
be tested, in regard both to phase boundaries and dynamics; and including for 
example the predicted universal scaling of SC spectra as the SC$\rightarrow$LM 
transition is approached --- leading to an $r$-dependent family of universal 
spectra, of which that well known to arise for the normal Anderson model should represent 
but a particular example, symptomatic of generic behaviour characteristic of the SC (or 
`generalized Fermi liquid') phase. These issues will be taken up in a subsequent 
publication [31], where predictions arising from the LMA will be shown to be 
remarkably well supported by benchmark NRG calculations. 
 
\ack
We are indebted to the many people with whom we have had stimulating discussions regarding the present work, with particular thanks to R. Bulla and T. Pruschke.  MTG acknowledges an EPSRC studentship, and we are further grateful to the British Council for financial support.
\clearpage
%\appendix
%\Alph{appendix}
%\Alph{section}
%\setcounter{section}{1}
\app
\section*{Appendix}

We outline the asymptotic behaviour of the self-consistency equation (4.6) for the mean-field local moment $|\mu|$, as $x=\frac{1}{2}U|\mu|\ra 0$; the SC/LM phase boundary at mean-field level can thereby be found, see section 4.1.  The general case of a finite hybridization/host bandwidth $D$ is considered, whence (section 4.1) the $D^0_\sigma(\w)$ contain pole contributions from outside the band ($|\w|>D$) for all $x \geq 0$; the corresponding poleweights are denoted by $Q^{\pm}_\sigma$ with $+/-$ for $\w>D$ and $\w<-D$ respectively.

From equation (4.6) the UHF self-consistency equation is thus
\appalpheqn
\bea
|\mu|&=\int^0_{-D}\mbox{d}\w\ \left[D^0_\up(\w)-D^0_\down(\w)\right]+\left[Q^{(-)}_\up-Q^{(-)}_\down\right]\\ 
&=f_{\band}(x)+\left[Q^{(-)}_\up-Q^{(-)}_\down\right]\\ 
&\equiv f(x)
\eea
\appreseteqn
where (from equation (4.3))
\be
\fl\ \  f_{\band}(x)=\int^0_{-D}\frac{\mbox{d}\w}{\pi}\ \Delta_{{\subi}}(\w)\left[\frac{1}{(\w+x-\Delta_{{\subr}}(\w))^2+\Delta_{{\subi}}^2(\w)}-\frac{1}{(\w-x-\Delta_{{\subr}}(\w))^2+\Delta_{{\subi}}^2(\w)}\right]
\ee
It is straightforward to show that the pole contributions $\left[Q^{(-)}_\up-Q^{(-)}_\down\right]\sim \mbox{O}(x)$ as $x \ra 0$; we thus focus on $f_{\band}(x)$.

\subsection*{(a) $0 \leq r<\frac{1}{2}$}
From (A.2),
\be
\left(\frac{\partial f_{\band}(x)}{\partial x}\right)_0=-\frac{4}{\pi}\int^0_{-D}\mbox{d}\w \ \frac{\Delta_{\subi}(\w)[\w-\Delta_{\subr}(\w)]}{\left([\w-\Delta_{\subr}(\w)]^2+\Delta_{\subi}^2(\w)\right)^2}.
\ee
The low-$\w$ behaviour of the integrand in (A.3) is $\sim |\w|^{-2r}$, whence the integral converges for $r<\frac{1}{2}$.  In consequence $f_{\band}(x)$, and hence $f(x)$, is $\mbox{O}(x)$ as $x \ra 0$; i.e. the exponent in equation (4.9) is $m=1$.  For $r<\frac{1}{2}$ the mean-field $U_{\cc} \equiv U^0_{\cc}(r)$ is thus finite and given by equation (4.10a).

\subsection*{(b) $\frac{1}{2}<r<1$}
The integral in (A.2) is controlled by its low-$\w$ behaviour, and we replace $
\Delta_{\subr}(\w)$ therein by its low-$\w$ form $\Delta_{\subr}(\w)\sim -\mbox{sgn}(\w)\beta(r)\Delta_0|\w|^r$ (equation (2.9)), where $\beta(r)=\mbox{tan}(\frac{\pi}{2}r)$.  Transforming the integration variable in (A.2) from $\w$ to $z=(\Delta_0/x)^{1/r}|\w|$ then yields for $r<1$
\be
\fl\ \ f_{\band}(x)\stackrel{x \ra 0}{\sim}x^{\frac{1-r}{r}}\frac{1}{\pi \Delta_0^{\frac{1}{r}}}\int^\infty_0\mbox{d}z\ z^r\left[\frac{1}{(1-\beta(r)z^r)^2+z^{2r}}-\frac{1}{(1+\beta(r)z^r)^2+z^{2r}}\right].
\ee
This integral converges only for $r>\frac{1}{2}$ whence, for $\frac{1}{2}<r<1$, $f_{\band}(x)$ and hence $f(x)$ have the $x \ra 0$ behaviour $f(x) \sim x^m$ with exponent $m=(1-r)/r <1$.  In consequence, $|\mu|$ vanishes only as $U \ra 0$, see equation (4.12).

\subsection*{(c) $r>1$}
From the normalization condition upon $D^0_\sigma(\w)$,
\be
\int^0_{-D}\mbox{d}\w\ D^0_\up(\w)=1-\int^D_0\mbox{d}\w\ D^0_\up(\w)-[Q^{(-)}_\up+Q^{(+)}_\up].
\ee
Hence from (A.1a,c) the full $f(x)$ is given by 
\be
f(x)=1-2\int^D_0\mbox{d}\w\ D^0_\up(\w)-[Q^{(+)}_\up+Q^{(-)}_\down]
\ee
where particle-hole symmetry is used ($D^0_\down(\w)=D^0_\up(-\w)$.  But as $x \ra 0$, $\int^D_0\mbox{d}\w \ D^0_\up(\w)$ reduces to $\int^D_0\mbox{d}\w \ d^{\band}_0(\w)$ (with $d^{\band}_0(\w)$ the band contribution to the non-interacting spectrum, see section 2.1); and for $x= 0$, $Q^{(+)}_\sigma=Q^{(-)}_\sigma \equiv Q_0$ is independent of $\sigma$.  Hence
\be
f(x=0)=1-2\int^D_0\mbox{d}\w\ d^{\band}_0(\w)-2Q_0.
\ee
But from normalization of the non-interacting spectrum $d_0(\w)$ (see equation (2.10a)),
\be
1=q+2\int^D_0\mbox{d}\w\ d^{\band}_0(\w)+2Q_0
\ee
where the weight, q, of the $\w=0$ pole in $d_0(\w)$ is given by equation (2.10b).  Hence, for $r>1$, $f(x=0)=q$; the exponent $m$ in equation (4.9a) is thus $m=0$, and in consequence the local moment
\be
|\mu|\stackrel{x \ra 0}{\sim}q
\ee
as follows from equation (4.8).

%\end{spacing}
\end{document}